\documentclass{article}
\usepackage{arxiv}
\usepackage{bm}
\usepackage[utf8]{inputenc} 
\usepackage[T1]{fontenc}    
\usepackage{hyperref}  
\usepackage{url}          
\usepackage{booktabs}     
\usepackage{amsfonts}      
\usepackage{nicefrac}  
\usepackage{microtype}     	
\usepackage{amsmath,amssymb,amsthm}
\usepackage{graphicx}
\usepackage[numbers]{natbib}
\usepackage{doi}
\usepackage{lineno}
\usepackage{physics}
\usepackage{color}
\usepackage[dvipsnames]{xcolor}
\usepackage{caption}
\usepackage{subcaption}
\usepackage{xargs}
\usepackage{algorithm}
\usepackage{algpseudocode}
\usepackage{xcolor}
\usepackage{physics}
\usepackage{color}
\usepackage[dvipsnames]{xcolor}
\usepackage{multirow}

\title{Physics-based deep kernel learning for parameter estimation in high dimensional PDEs}

\author{
        \href{https://orcid.org/0009-0003-3309-324X}{\includegraphics[scale=0.06]{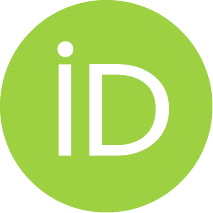}\hspace{1mm}Weihao ~Yan*} \\
	Mathematics of Imaging \& AI\\
	University of Twente, The Netherlands \\
	\texttt{w.yan@utwente.nl} \\
        \And
	\href{https://orcid.org/0000-0003-0145-5069}{\includegraphics[scale=0.06]{orcid.pdf}\hspace{1mm}Christoph ~Brune} \\
	Mathematics of Imaging \& AI\\
	University of Twente, The Netherlands \\
	\texttt{c.brune@utwente.nl} \\
	\And
	\href{https://orcid.org/0000-0002-5541-437X}{\includegraphics[scale=0.06]{orcid.pdf}\hspace{1mm}Mengwu ~Guo} \\
	Centre for Mathematical Sciences\\
	Lund University, Sweden \\
	\texttt{mengwu.guo@math.lu.se} \\
}

\date{}

\begin{document}
\maketitle

\begin{abstract}
Inferring parameters of high-dimensional partial differential equations (PDEs) poses significant computational and inferential challenges, primarily due to the curse of dimensionality and the inherent limitations of traditional numerical methods. This paper introduces a novel two-stage Bayesian framework that synergistically integrates  training, physics-based deep kernel learning (DKL) with Hamiltonian Monte Carlo (HMC) to robustly infer unknown PDE parameters and quantify their uncertainties from sparse, exact observations. The first stage leverages physics-based DKL to train a surrogate model, which jointly yields an optimized neural network feature extractor and robust initial estimates for the PDE parameters. In the second stage, with the neural network weights fixed, HMC is employed within a full Bayesian framework to efficiently sample the joint posterior distribution of the kernel hyperparameters and the PDE parameters. Numerical experiments on canonical and high-dimensional inverse PDE problems demonstrate that our framework accurately estimates parameters, provides reliable uncertainty estimates, and effectively addresses challenges of data sparsity and model complexity, offering a robust and scalable tool for diverse scientific and engineering applications.
\end{abstract}


\section{Introduction}
In various computational scientific and engineering fields, deterministic numerical methods such as finite element and finite difference are widely used to solve PDEs. 
These methods approximate solutions through numerical discretization, which inevitably introduces errors and uncertainties.
Moreover, inverse problems require repeatedly employing forward solvers to infer unknown parameters from observed data \cite{chkrebtii2016bayesian}. 
As a result, in inverse problems, uncertainty quantification (UQ) \cite{roy2011comprehensive} becomes both critical and challenging, and deterministic methods often struggle to handle these uncertainties effectively.
To address this issue, \textit{probabilistic numerical methods} have emerged \cite{hennig2015probabilistic}. For example, \cite{conrad2017statistical} proposed an approach that constructs a global uncertainty distribution based on local Gaussian assumptions, thus providing a statistical interpretation of numerical methods. 
The Bayesian method offers a natural solution, demonstrating an advantage in solving inverse problems by systematically quantifying uncertainties based on observed data and prior knowledge \cite{stuart2010inverse, bilionis2013solution}. 
Building on this, \textit{Bayesian probabilistic numerical methods} \cite{Cockayne2019Bayesian, owhadi2015bayesian} provide an even more systematic and rigorous framework for quantifying uncertainties.
However, most approaches often rely on Monte Carlo sampling, which is extremely expensive because they typically require even tens of thousands of evaluations of simulations \cite{bui2012extreme}. 

Therefore, surrogate models are commonly needed. In particular, within a Bayesian framework, Gaussian process (GP) \cite{williams2006gaussian}  surrogate models have been effectively employed to solve PDE problems \cite{raissi2017machine, raissi2018numerical, raissi2018hidden} by providing a probabilistic surrogate that quantifies the epistemic uncertainty, often outperforming Monte Carlo methods for problems with relatively low-dimensional random inputs \cite{bilionis2013solution}.
However, GP methods face challenges in high-dimensional settings because the number of required samples typically grows exponentially with the input dimension, which is known as the \textit{curse of dimensionality} \cite{bengio2005curse}.
The concept of the \textit{active subspace} (AS) has therefore been introduced. AS is defined as a low-dimensional linear manifold within the high-dimensional input space along which the output exhibits the most variation \cite{russi2010uncertainty, constantine2014active}. 
However, many physical models are too complex to be captured by a strictly AS; a nonlinear low-dimensional structure may be hidden within the high-dimensional space. 

\textit{Deep kernel learning} has emerged as a powerful solution, as it synergistically combines the ability of deep neural networks to learn nonlinear feature representations with the probabilistic modelling of GPs \cite{wilson2016deep, calandra2016manifold}. 
By learning an explicit mapping from the high-dimensional input space to a low-dimensional feature space where the GP kernel operates, DKL is exceptionally well-suited to uncovering these hidden nonlinear structures. This paper leverages DKL as a core component to address parameter estimation in high dimensional PDEs. 
While DKL presents an ideal candidate for the surrogate model, a critical challenge remains: how to robustly and efficiently integrate this powerful surrogate into a physics-based Bayesian inference workflow? A naive approach that attempts to simultaneously train the high-capacity neural network, tune the GP kernel, and infer the physical parameters often leads to a complex, non-convex optimization landscape, resulting in instability and sensitivity to initialization.

This paper addresses this methodological challenge by introducing a novel two-stage Bayesian framework that separates the surrogate model's representation learning from the final parameter inference. Our approach synergistically integrates physics-informed training, physics-based DKL, and HMC to achieve robust and efficient inference for high-dimensional inverse problems. We will demonstrate that this structured approach not only mitigates the optimization difficulties inherent in one frameworks but also leads to more accurate and reliable parameter estimates, even when dealing with sparse observational data.

\subsection{Related work}
The computational challenges of high-dimensional inverse problems have motivated diverse methodological developments at the intersection of numerical analysis and machine learning. Traditional Bayesian approaches often rely on GP surrogates to emulate expensive forward solvers \cite{kennedy2001bayesian}, frequently combined with dimension reduction techniques like active subspaces to manage moderate-dimensional parameter spaces \cite{constantine2014active}. While effective, these methods often treat the forward solver as a black box. The advent of physics-informed machine learning, particularly through automatic differentiation, has opened new pathways to embed physical laws more deeply into the learning process.

This field has largely evolved along two  paths. The first is the physics-informed neural network (PINN), which incorporates the PDE residual as a soft constraint in the network's loss function \cite{raissi2017machine, raissi2018numerical,raissi2018hidden}. While PINNs can solve inverse problems by making the parameters trainable, standard PINNs provide only point estimates. Bayesian PINN variants aim to provide uncertainty quantification but require inference over the high-dimensional weight space of the network, which is computationally demanding and can be difficult to scale \cite{yang2021b}.

The second path, which is more aligned with our work, involves encoding physical laws directly into the structure of a GP prior. This mathematically rigorous approach from early work on encoding linear operators into kernels \cite{raissi2017machine} into a comprehensive theory of probabilistic numerics \cite{Cockayne2019Bayesian}. Recent breakthroughs have established that such PDE-constrained GPs can generalize classical linear PDE solvers like the finite element method, providing a unified framework for quantifying discretization errors and model mismatch \cite{pfortner2022physics, girolami2021statistical}. This paradigm has been successfully extended to solve complex nonlinear PDEs and their associated inverse problems \cite{chen2021solving, li2024parameter}, and even to enforce fundamental conservation laws by structuring the GP prior according to port-Hamiltonian systems \cite{beckers2022gaussian}. However, this powerful approach faces two key challenges when applied to high-dimensional inverse problems. First, since these methods are built on a standard GP framework, they are limited by the curse of dimensionality in high-dimensional parameter spaces. Second, the combination of the GP kernel hyperparameters and the unknown PDE parameters results in a complex, non-convex optimization landscape, making the inference process highly sensitive to initialization and potentially unstable. This highlights a critical research gap: an ideal framework must combine the principled physical constraints of these GP methods with the capability to handle high-dimensional spaces, while also addressing the resulting optimization complexity. Our work addresses this gap by proposing a physics-informed DKL approach that uses a novel two-stage framework, designed specifically to overcome both of these challenges.

Moreover, while PDE-constrained GPs focus on embedding physics in the prior, DKL  has been developed primarily as a powerful data-driven tool for high-dimensional regression, demonstrating excellent performance in learning nonlinear data representations \cite{wilson2016deep}. Recent work has begun to adapt PDE- constrained DKL architectures for solving forward PDE problems \cite{yan2025pde}, and its core idea of combining NNs with GPs is central to the broader field of operator learning \cite{mora2025operator}. This establishes DKL as a state-of-the-art component for building a surrogate model that can handle high-dimensional parameter spaces. However, simply selecting DKL as the surrogate does not resolve the challenge of how to best integrate it into a robust inverse problem workflow. For the inference stage itself, HMC has become the gold standard for sampling from complex, high-dimensional distributions due to its efficiency in exploring the parameter space \cite{neal2011mcmc}. The performance of HMC, however, is highly dependent on its initialization, and its efficiency can be severely restricted without a well-informed starting point in a region of high posterior probability. Despite the individual advances in these areas, a framework that synergistically combines the representation power of DKL, the rigour of physics-informed constraints, and the sampling efficiency of HMC in a computationally robust and stable manner remains an open challenge. This work aims to fill that gap by introducing a structured, two-stage approach that separates the complex task of representation learning from the final parameter inference, thereby addressing both the optimization of the surrogate and the efficiency of the Bayesian computation.


\section{Contributions}
Our work introduces a novel framework for parameter estimation in systems governed by high-dimensional PDEs, leveraging a physics-based DKL approach. The primary contributions of this work are as follows:

\begin{enumerate}
    \item We propose and develop a structured, two-stage framework for parameter estimation in high-dimensional PDE systems. This framework is explicitly designed to overcome the optimization and stability challenges inherent in single approaches that attempt to simultaneously train a surrogate model and estimate physical parameters. By separating the complex task of representation learning from the final parameter estimation, our approach significantly enhances the robustness and efficiency of the entire workflow.

    \item The first stage of our framework consists of a physics-informed pre-training procedure. This stage serves a dual purpose: (i) it trains the neural network component of the DKL model to be a highly accurate surrogate for the PDE solution by minimizing a composite loss function that includes both data fidelity and PDE residual terms; and (ii) it provides a high-quality point estimate for the unknown physical parameters, which serves as a well-informed starting point for the subsequent Bayesian inference.

    \item The second stage leverages the optimized surrogate from the first stage to perform efficient and rigorous uncertainty quantification for the estimated parameters. With the expressive neural network weights held fixed, we employ HMC to sample from the low-dimensional joint posterior distribution of the PDE parameters and the remaining GP kernel hyperparameters. This strategy makes the Bayesian inference tractable, avoiding the prohibitive cost of sampling in the high-dimensional weight space of the neural network, and yields reliable uncertainty estimates for the inferred parameters even in data-sparse scenarios.
\end{enumerate}

\subsection{Structure of the paper}
The remainder of this paper is organized as follows. Section \ref{sec:background} reviews foundational concepts in GP, deep kernel learning, physics-based machine learning approaches, the Bayesian framework for inverse problems, and HMC. Section \ref{sec:method} details our proposed two-stage methodology: it begins with the problem formulation, then describes the development of the physics-based  DKL surrogate, the physics-based pretraining (stage 1) for model and parameter initialization, and finally the Bayesian inference (stage 2) for PDE parameter estimation and uncertainty quantification using HMC. In section \ref{sec:numerical_experiments}, we demonstrate the efficacy and robustness of our framework through numerical experiments on inverse problems involving high-dimensional PDEs. Section \ref{sec:discussion} discusses the findings and limitations of our approach. Finally, Section \ref{sec:conclusion_outlook} concludes the paper and outlines future research directions. Some mathematical derivations are provided in Appendix \ref{app:likelihood_derivation}.

\section{Background}
\label{sec:background}
\subsection{Gaussian processes regression and deep kernel learning} \label{gp_dkl}
\paragraph{Gaussian process regression}
Gaussian process regression (GPR) is a non-parametric Bayesian approach for regression, which offers a principled method to quantify uncertainty in predictions. A GP is a collection of random variables, any finite number of which have a joint Gaussian distribution. It defines a prior distribution over functions. 

Let $f: \mathcal{S} \to \mathbb{R}$ be a latent function, where $\mathcal{S} \subseteq \mathbb{R}^d$ is the $d$-dimensional input domain. The function $f$ is modelled by a GP prior:
\begin{equation} \label{eq:gp_prior}
f(\mathbf{s}) \sim \mathcal{GP}\big(\mu(\mathbf{s}),\, \kappa(\mathbf{s},\mathbf{s}')\big)\,,
\end{equation}
where $\mathbf{s}, \mathbf{s}' \in \mathcal{S}$. The function $\mu: \mathcal{S} \to \mathbb{R}$ is the mean function, and $\kappa: \mathcal{S} \times \mathcal{S} \to \mathbb{R}$ is the positive semi-definite covariance function, also known as the kernel. For notational simplicity, the prior mean function $\mu(\mathbf{s})$ is often assumed to be zero, i.e., $\mu(\mathbf{s}) \equiv 0$.

Given a training dataset $\mathcal{D} = \{(\mathbf{s}_i, y_i)\}_{i=1}^n$, consisting of $n$ input-output pairs where $\mathbf{s}_i \in \mathcal{S}$ and $y_i \in \mathbb{R}$, the observations $y_i$ are related to the latent function $f$ through the model:
\begin{equation} \label{eq:gpr_observation_model}
y_i = f(\mathbf{s}_i) + \varepsilon_i\,,
\end{equation}
where $\varepsilon_i$ are independent and identically distributed (i.i.d.) Gaussian noise terms, $\varepsilon_i \sim \mathcal{N}(0, \tau^2)$, with $\tau^2$ being the noise variance.

Let $\mathbf{S} = [\mathbf{s}_1, \ldots, \mathbf{s}_n]^T \in \mathbb{R}^{n \times d}$ denote the matrix of training inputs and $\mathbf{y} = [y_1, \ldots, y_n]^T \in \mathbb{R}^n$ be the vector of corresponding training outputs. For a new test input $\mathbf{s}_* \in \mathcal{S}$, the predictive distribution of the latent function value $f(\mathbf{s}_*)$, conditioned on the training data $\mathcal{D}$  is Gaussian: $f(\mathbf{s}_*) \mid \mathcal{D}, \mathbf{s}_* \sim \mathcal{N}(\mu_*(\mathbf{s}_*), \sigma_*^2(\mathbf{s}_*))$. The predictive mean $\mu_*(\mathbf{s}_*)$ and variance $\sigma_*^2(\mathbf{s}_*)$ are given by:
\begin{align}
\mu_*(\mathbf{s}_*) &= \mu(\mathbf{s}_*) + \mathbf{k}_{*(\mathbf{s}_*)}^T (\mathbf{K}_{\mathbf{S}\mathbf{S}}+\tau^2 I_n)^{-1} (\mathbf{y} - \boldsymbol{\mu}_{\mathbf{S}})\,, \label{eq:gpr_predictive_mean} \\
\sigma_*^2(\mathbf{s}_*) &= \kappa(\mathbf{s}_*,\mathbf{s}_*) - \mathbf{k}_{*(\mathbf{s}_*)}^T (\mathbf{K}_{\mathbf{S}\mathbf{S}}+\tau^2 I_n)^{-1} \mathbf{k}_{*(\mathbf{s}_*)}\,. \label{eq:gpr_predictive_variance}
\end{align}
In these expressions, $\boldsymbol{\mu}_{\mathbf{S}} = [\mu(\mathbf{s}_1), \ldots, \mu(\mathbf{s}_n)]^T \in \mathbb{R}^n$ is the prior mean vector evaluated at the training inputs. $\mathbf{K}_{\mathbf{S}\mathbf{S}} \in \mathbb{R}^{n \times n}$ is the covariance matrix with elements $[\mathbf{K}_{\mathbf{S}\mathbf{S}}]_{ij} = \kappa(\mathbf{s}_i, \mathbf{s}_j)$. The vector $\mathbf{k}_{*(\mathbf{s}_*)} = [\kappa(\mathbf{s}_1,\mathbf{s}_*), \ldots, \kappa(\mathbf{s}_n,\mathbf{s}_*)]^T \in \mathbb{R}^{n \times 1}$ contains the covariances between the $n$ training inputs and the test input $\mathbf{s}_*$. $\kappa(\mathbf{s}_*,\mathbf{s}_*)$ is the prior variance at $\mathbf{s}_*$, and $I_n$ is the $n \times n$ identity matrix.

\paragraph{Deep kernel learning}
DKL \cite{wilson2016deep} enhances the expressive capacity of GPs by parameterizing the kernel function using a DNN. This synergistically integrates the representation learning power of DNNs with the probabilistic framework of GPs.

The core idea of DKL is to transform inputs $\mathbf{s} \in \mathcal{S}$ via a neural network embedding function $g_{\boldsymbol{\theta}}: \mathcal{S} \to \mathbb{R}^p$, parameterized by weights $\boldsymbol{\theta}$, into a $p$-dimensional feature space. A base kernel $\kappa_{\text{base}}$ then operates on these features. The deep kernel $\kappa_{\text{DKL}}$ is:
\begin{equation} \label{eq:dkl_kernel_definition}
\kappa_{\text{DKL}}(\mathbf{s}, \mathbf{s}'; \boldsymbol{\theta}, \boldsymbol{\psi}) = \kappa_{\text{base}}\big(g_{\boldsymbol{\theta}}(\mathbf{s}), g_{\boldsymbol{\theta}}(\mathbf{s}'); \boldsymbol{\psi}\big)\,,
\end{equation}
where $\kappa_{\text{base}}$ has hyperparameters $\boldsymbol{\psi}$. The DKL model parameters are thus $(\boldsymbol{\theta}, \boldsymbol{\psi})$. These, along with noise variance $\tau^2$, are typically learned by maximizing the log marginal likelihood of observations $\mathbf{y}$:
\begin{equation} \label{eq:dkl_log_marginal_likelihood}
\log p(\mathbf{y} \mid \mathbf{S}, \boldsymbol{\theta}, \boldsymbol{\psi}, \tau^2) = -\frac{1}{2}(\mathbf{y}-\boldsymbol{\mu}_{\mathbf{S}})^T (\mathbf{K}_{{\text{DKL}},{\mathbf{S}\mathbf{S}}} + \tau^2 I_n)^{-1} (\mathbf{y}-\boldsymbol{\mu}_{\mathbf{S}}) - \frac{1}{2}\log|\mathbf{K}_{{\text{DKL}},{\mathbf{S}\mathbf{S}}} + \tau^2 I_n| - \frac{n}{2}\log(2\pi)\,.
\end{equation}
Here, $\mathbf{K}_{{\text{DKL}},{\mathbf{S}\mathbf{S}}} \in \mathbb{R}^{n \times n}$ is the kernel matrix evaluated at the training inputs $\mathbf{S}$ using the deep kernel, i.e., $[\mathbf{K}_{\text{DKL}}]_{ij} = \kappa_{\text{DKL}}(\mathbf{s}_i, \mathbf{s}_j; \boldsymbol{\theta}, \boldsymbol{\psi})$. The prior mean vector $\boldsymbol{\mu}_{\mathbf{S}} \in \mathbb{R}^n$ denotes the mean of the GP evaluated at $\mathbf{S}$. The parameters with a superscript ${ }^*$, i.e., $\left(\boldsymbol{\theta}^*, \boldsymbol{\psi}^*\right)$ and $\tau^{2 *}$, denote the optimized values obtained by maximizing the marginal likelihood.

Once the optimal parameters $(\boldsymbol{\theta}^*, \boldsymbol{\psi}^*)$ and noise variance $\tau^{2*}$ have been learned, the DKL model is used for prediction. A key insight is that the trained DKL surrogate operates as a standard GP whose covariance is given by the learned deep kernel. Therefore, predictions for a new test input $\mathbf{s}_*$ are made using the established GPR predictive equations (Eqs. \eqref{eq:gpr_predictive_mean} and \eqref{eq:gpr_predictive_variance}). This is achieved by replacing the base kernel $\kappa(\cdot, \cdot)$ with the optimized deep kernel, $\kappa_{\text{DKL}}(\cdot, \cdot; \boldsymbol{\theta}^*, \boldsymbol{\psi}^*)$, and using the optimized noise variance $\tau^{2*}$. Operationally, this means the kernel matrix $\mathbf{K}_{\mathbf{S}\mathbf{S}}$ and covariance vector $\mathbf{k}_{*(\mathbf{s}_*)}$ are first computed using the learned deep kernel before being used in the predictive equations to obtain the final mean and variance.

\subsection{Machine learning frameworks for physics-based  systems} \label{sec:ml_pde_constrained}
Embedding physical laws, typically formulated as PDEs, within machine learning models is a key strategy for developing robust and data-efficient surrogates. This section reviews two approaches: physics-informed neural networks (PINNs) and PDE-constrained Gaussian processes.

\paragraph{Physics-informed neural networks}
PINNs utilize a neural network $u_{\theta}(\mathbf{s}; \boldsymbol{\theta}_{\text{NN}})$ to construct a differentiable surrogate for the solution of a given PDE system \cite{raissi2019physics}. Here, $\mathbf{s}$ denotes spatio-temporal coordinates, and $\boldsymbol{\theta}_{\text{NN}}$ are the learnable parameters of the network. The training of $u_{\theta}$ is guided by minimizing a composite loss function. For a PDE system characterized by a differential operator $\mathcal{A}$ and source term $f(\mathbf{s})$, such that the governing equation is $\mathcal{A}[u(\mathbf{s})] = f(\mathbf{s})$ in a domain $\Omega$, and subject to boundary/initial conditions $\mathcal{B}[u(\mathbf{s})] = h(\mathbf{s})$ on $\partial\Omega$, the PINN loss function is typically structured as:
\begin{equation} \label{eq:pinn_loss_u_solve_f_src}
\mathcal{L}_{\text{PINN}}(\boldsymbol{\theta}_{\text{NN}}) =  w_{\text{PDE}} \mathcal{L}_{\text{PDE}}(\boldsymbol{\theta}_{\text{NN}}) + w_{\text{bc}} \mathcal{L}_{\text{bc}}(\boldsymbol{\theta}_{\text{NN}}) \,.
\end{equation}
The PDE residual term $\mathcal{L}_{\text{PDE}}$, enforces the governing physical law by penalizing deviations from it at $N_{\text{col}}$ collocation points, $\{\mathbf{s}_j^{\text{col}}\}$, sampled within $\Omega$:
\begin{equation} \label{eq:pinn_loss_pde_u_solve_f_src}
\mathcal{L}_{\text{PDE}}(\boldsymbol{\theta}_{\text{NN}}) = \frac{1}{N_{\text{col}}} \sum_{j=1}^{N_{\text{col}}} \| \mathcal{A}[u_{\theta}(\mathbf{s}_j^{\text{col}}; \boldsymbol{\theta}_{\text{NN}})] - f(\mathbf{s}_j^{\text{col}}) \|^2 \,, 
\end{equation}
and 
\begin{equation}
\mathcal{L}_{\text{bc}}(\boldsymbol{\theta}_{\text{NN}})
= \frac{1}{N_{\text{bc}}}
  \sum_{k=1}^{N_{\text{bc}}}
  \bigl\|\mathcal{B}\bigl[u_{\theta}(\mathbf{s}_k^{\text{bc}};\boldsymbol{\theta}_{\text{NN}})\bigr]
    - h(\mathbf{s}_k^{\text{bc}})\bigr\|^2 \,,
\end{equation}
enforces the boundary/initial conditions on $N_{\text{bc}}$ points $\{\mathbf{s}_k^{\text{bc}}\}\in \partial\Omega$. Automatic differentiation (AD) is essential for computing the derivatives of $u_{\theta}$ with respect to $\mathbf{s}$ as required by the operators $\mathcal{A}$ and $\mathcal{B}$. The non-negative weights $w_{\text{PDE}}, w_{\text{bc}}$ balance the influence of each component. Minimizing $\mathcal{L}_{\text{PINN}}$ with respect to $\boldsymbol{\theta}_{\text{NN}}$ yields a neural network surrogate $u_{\theta}$ that approximates the PDE solution by fitting the data while enforcing the physical constraints.

\paragraph{PDE-constrained Gaussian processes}
\label{par:pde_constrained_gps}
GPs offer a probabilistic framework for incorporating constraints from linear PDEs by exploiting their behaviour under linear transformations. If a function $u(\mathbf{s})$ is modelled as a GP, $u \sim \mathcal{GP}(m_u(\mathbf{s}), k_{uu}(\mathbf{s}, \mathbf{s}'))$, then for any linear differential operator $\mathcal{A}$, the transformed function $f(\mathbf{s}) = \mathcal{A}u(\mathbf{s})$ is also a GP. 

The mean of this transformed GP is $m_f(\mathbf{s}) = \mathcal{A}m_u(\mathbf{s})$. The covariance functions involving $f$ are derived by applying the operator $\mathcal{A}$ to the original covariance function $k_{uu}$; specifically, the autocovariance of $f$ is $k_{ff}(\mathbf{s}, \mathbf{s}') = \mathcal{A}_{\mathbf{s}}\mathcal{A}_{\mathbf{s}'}k_{uu}(\mathbf{s}, \mathbf{s}')$, and the cross-covariance between $u$ and $f$ is $k_{uf}(\mathbf{s}, \mathbf{s}') = \mathcal{A}_{\mathbf{s}'}k_{uu}(\mathbf{s}, \mathbf{s}')$, where $\mathcal{A}_{\mathbf{s}}$ and $\mathcal{A}_{\mathbf{s}'}$ denote the operator $\mathcal{A}$ acting on the first and second arguments of the kernel function, respectively.

This property enables the direct incorporation of a linear PDE written as $\mathcal{A}[u(\mathbf{s}); \boldsymbol{\phi}] = f(\mathbf{s})$  into the GP model. By defining a joint GP over $u$ and $f$ (as $f = \mathcal{A}u$), observations of both the solution $u$ and the forcing term $f$ can be assimilated through a joint likelihood. 

\subsection{The Bayesian approach to parameter estimation} \label{sec:bayesian_inverse_problems}
The Bayesian scheme provides the theoretical foundation for inferring unknown parameters $\boldsymbol{\phi}$ from observed data $\mathbf{d}$. This inference is achieved by characterizing the parameters through a posterior probability distribution $p(\boldsymbol{\phi} \mid \mathbf{d})$, which is obtained via Bayes' theorem:
\begin{equation} \label{eq:bayes_theorem}
p(\boldsymbol{\phi} \mid \mathbf{d}) \propto p(\mathbf{d} \mid \boldsymbol{\phi}) p(\boldsymbol{\phi}) \,.
\end{equation}
The construction of the posterior relies on two components: the prior distribution $p(\boldsymbol{\phi})$, which encodes pre-existing knowledge and/or imposes regularization on $\boldsymbol{\phi}$, and the likelihood function $p(\mathbf{d} \mid \boldsymbol{\phi})$. The likelihood is determined by a forward model $\mathcal{G}(\boldsymbol{\phi})$, mapping parameters to the observables, and a statistical model for the observation noise; for instance, an additive Gaussian noise with covariance $\boldsymbol{\Sigma}_{\text{noise}}$ yields $p(\mathbf{d} \mid \boldsymbol{\phi}) \propto \exp(-\frac{1}{2} ||\mathbf{d} - \mathcal{G}(\boldsymbol{\phi})||^2_{\boldsymbol{\Sigma}_{\text{noise}}^{-1}})$. The posterior $p(\boldsymbol{\phi} \mid \mathbf{d})$ thus integrates all available information regarding $\boldsymbol{\phi}$.

The objective of Bayesian inference is the derivation of $p(\boldsymbol{\phi} \mid \mathbf{d})$. However, for problems where the forward model $\mathcal{G}(\boldsymbol{\phi})$ approximates the solution to a PDE defined on a high-dimensional domain, the posterior distribution is typically analytically intractable and the computational cost of inference can be prohibitive. This requires employing numerical approximation techniques, such as Markov Chain Monte Carlo (MCMC) methods, to generate samples from $p(\boldsymbol{\phi} \mid \mathbf{d})$ for estimating its properties.

\subsection{Hamiltonian Monte Carlo} \label{sec:hmc}
HMC \cite{neal2011mcmc} is an MCMC algorithm designed for efficient sampling from continuous probability distributions, particularly those defined over high-dimensional spaces where traditional MCMC methods may suffer from slow convergence due to random walk behavior. HMC achieves this by leveraging Hamiltonian dynamics to propose distant states with high acceptance probabilities. The target parameters $\boldsymbol{\phi}$ are augmented with auxiliary momentum variables $\mathbf{p}$, forming a phase space $(\boldsymbol{\phi}, \mathbf{p})$. A Hamiltonian function $H(\boldsymbol{\phi}, \mathbf{p})$ is defined as:
\begin{equation} \label{eq:hamiltonian}
H(\boldsymbol{\phi}, \mathbf{p}) = U(\boldsymbol{\phi}) + V(\mathbf{p}) \,,
\end{equation}
where the potential energy $U(\boldsymbol{\phi}) = -\log [p(\mathbf{d} \mid \boldsymbol{\phi})p(\boldsymbol{\phi})]$ is the negative logarithm of the unnormalized target posterior density for $\boldsymbol{\phi}$ (Eq.\eqref{eq:bayes_theorem}), and $V(\mathbf{p}) = \frac{1}{2}\mathbf{p}^T M^{-1} \mathbf{p}$ is the kinetic energy, with $M$ being a symmetric positive-definite mass matrix.

To generate a proposal, HMC simulates the evolution of $(\boldsymbol{\phi}, \mathbf{p})$ according to Hamilton's equations of motion: 
\begin{equation}
\left\{
\begin{array}{l}
\dot{\boldsymbol{\phi}} = \nabla_{\mathbf{p}} V(\mathbf{p}) = M^{-1}\mathbf{p} \\
\dot{\mathbf{p}} = -\nabla_{\boldsymbol{\phi}} U(\boldsymbol{\phi}) \,.
\end{array}
\right .
\end{equation}
These dynamics theoretically conserve the total energy $H(\boldsymbol{\phi}, \mathbf{p})$. In practice, since the continuous-time Hamiltonian equations cannot be solved analytically, they are simulated for a finite duration using a numerical integrator. This process involves discretizing the trajectory into $L$ steps of a small step size $\epsilon$. The leapfrog method \cite{neal2011mcmc} is typically chosen for this task as it is both symplectic and time-reversible, properties that are crucial for maintaining stability and high acceptance rates. 
Starting from the current state $\boldsymbol{\phi}^{(k)}$ and a randomly drawn momentum $\mathbf{p}^{(k)} \sim \mathcal{N}(\mathbf{0}, M)$, the leapfrog integrator is applied for $L$ steps with a step size $\epsilon$ to reach a proposed state $(\boldsymbol{\phi}', \mathbf{p}')$. Due to discretization errors, this proposal is accepted with probability $
  \refstepcounter{equation}\label{eq:hmc_acceptance}
  \min\!\bigl(1,\exp[-H(\boldsymbol{\phi}',\mathbf{p}')
       + H(\boldsymbol{\phi}^{(k)},\mathbf{p}^{(k)})]\bigr)
$.
If accepted, $\boldsymbol{\phi}^{(k+1)} = \boldsymbol{\phi}'$; otherwise, $\boldsymbol{\phi}^{(k+1)} = \boldsymbol{\phi}^{(k)}$. The momentum $\mathbf{p}$ is typically resampled from its conditional distribution at each iteration. Under the joint target $\pi(\phi, \mathbf{p})$, the conditional momentum is independent of $\phi$ and equals $\mathbf{p}^{(k)}$. Hence we perform a full momentum refresh at each iteration by sampling $\mathbf{p}^{(k)} \sim \mathcal{N}(\mathbf{0}, M)$ independently of the past.  HMC's efficacy relies on the availability of the gradient $\nabla_{\boldsymbol{\phi}} U(\boldsymbol{\phi})$, i.e., the gradient of the log target density with respect to $\boldsymbol{\phi}$.

\section{Methodology}
\label{sec:method}
We propose a probabilistic framework that integrates physics-based  DKL with Bayesian inference to address the inverse problem of inferring unknown parameters $\vb*{\phi}$ embedded within the linear operators defining a time-dependent PDE. $\vb*{\phi}$ denotes the $d_\phi$-dimensional vector of unknown PDE parameters. The inference is conditioned upon a set of observations of both the system's state variable $u$ and its associated source term $f$. 

Our method proceeds in two primary stages: (i) physics-based pretraining of a DKL-based surrogate model for the solution field $u$, designed to yield effective initializations for $\vb*{\phi}$ and the DKL parameters and hyperparameters; and (ii) subsequent full Bayesian inference via HMC  to rigorously quantify the posterior uncertainty of the inferred PDE parameters $\vb*{\phi}$. The resulting posterior mean subsequently enables the prediction of the solution field $u(\vb*{x}, t)$, along with its associated uncertainties, by employing the trained physics-based  DKL model as a forward solver conditioned on the inferred parameters $\boldsymbol{\phi}$.

\begin{figure}[!ht]
    \centering
    \includegraphics[width=1\textwidth]{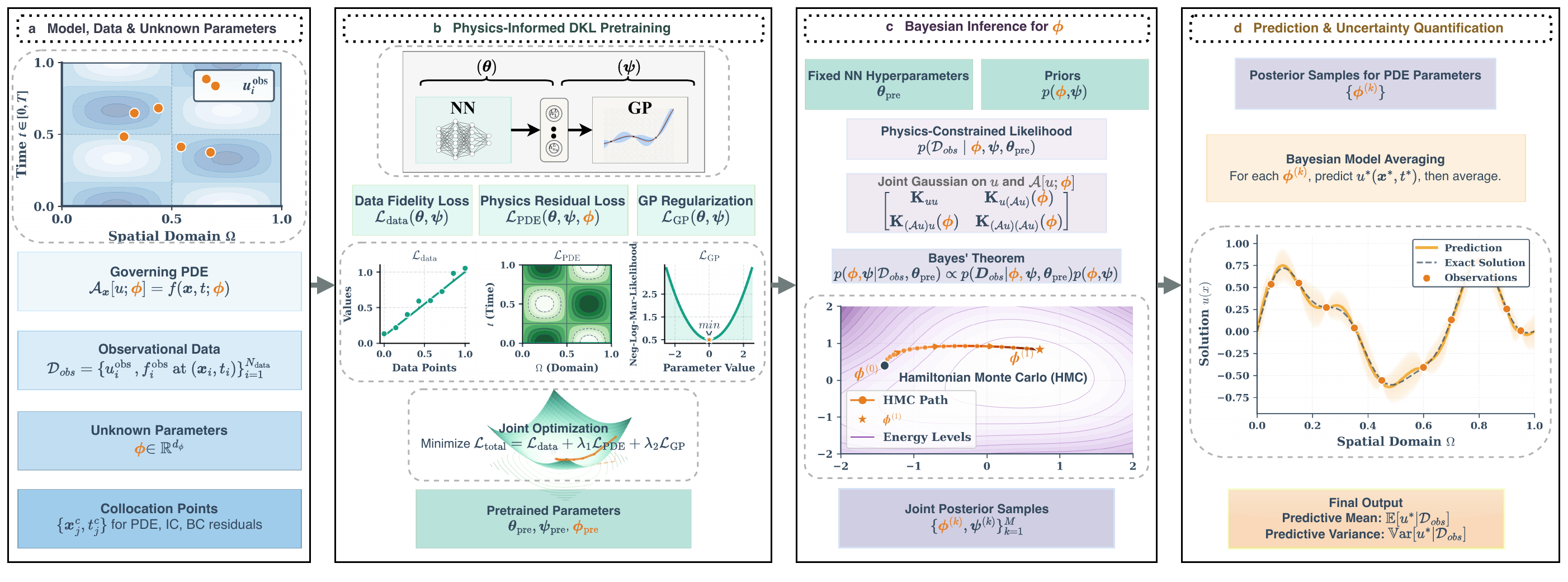}
    \caption{Schematic of the proposed two-stage Bayesian inference framework for PDE parameter estimation using physics-based  DKL. (a) Problem definition: inputs include the governing PDE with unknown parameters $\boldsymbol{\phi}$, and observational data $\mathcal{D}_{{\text{obs}}}$ (for state $u$ and source $f$). (b) (Stage 1) physics-based pretraining: a composite loss function is minimized to obtain optimized DKL neural network parameters $\boldsymbol{\theta}_{\text{pre}}$ and initial estimates for DKL kernel hyperparameters $\boldsymbol{\psi}_{\text{pre}}$ and PDE parameters $\boldsymbol{\phi}_{\text{pre}}$. (c) (Stage 2) Bayesian inference: with $\boldsymbol{\theta}$, HMC samples the joint posterior $p(\boldsymbol{\phi}, \boldsymbol{\psi} \mid \mathcal{D}_{{\text{obs}}}, \boldsymbol{\theta})$. (d) Outcome: parameter UQ and solution prediction: marginal posterior samples for $\boldsymbol{\phi}$ are derived for uncertainty quantification, which can then inform predictions of the solution field $u(\mathbf{s})$ with associated uncertainty.}
    \label{fig:overall_workflow}
\end{figure}

A schematic overview of this entire two-stage framework, illustrating its core components, is presented in Figure \ref{fig:overall_workflow}. Subsequent subsections will elaborate on each part of this workflow.

\subsection{Problem formulation}
\label{sec:problem_formulation}
We consider a time-dependent PDE governing the state variable $u(\vb*{x}, t)$, defined over a spatial domain $\Omega \subset \mathbb{R}^{d_x}$ and time interval $[0, T]$. The PDE is expressed as:
\begin{equation}
\partial_t u(\vb*{x}, t) = \mathcal{A}_{\vb*{x}}[u(\vb*{x}, t); \vb*{\phi}] + f(\vb*{x}, t; \vb*{\phi}), \quad (\vb*{x}, t) \in \Omega \times [0, T]\,,
\label{eq:pde_definition_method}
\end{equation}
subject to appropriate initial and boundary conditions. Here, $\mathcal{A}_{\vb*{x}}$ is a linear spatial differential operator and $f$ is a source term; both may depend on the unknown parameter vector $\vb*{\phi}$. For conciseness, we define the spatio-temporal linear operator $\mathcal{A}[u; \vb*{\phi}] := \partial_t u - \mathcal{A}_{\vb*{x}}[u; \vb*{\phi}]$, allowing Eq.~\eqref{eq:pde_definition_method} to be rewritten as: 
\begin{equation}
\mathcal{A}[u(\cdot); \vb*{\phi}] = f(\cdot; \vb*{\phi}) \,.
\label{eq:pde_operator_form_method}
\end{equation}
The inverse problem is to infer $\vb*{\phi}$ from a dataset of observations, $\mathcal{D}_{{\text{obs}}}$. This dataset consists of exact (noiseless) measurements. Specifically, it includes $N_{u}$ observations of the state variable $u$, denoted by a vector $\mathbf{u}_{{\text{obs}}} = \{u_i^{\text{obs}} = u(\vb*{s}_{\mathbf{u},i})\}_{i=1}^{N_{u}}$, recorded at spatio-temporal locations $\mathbf{S}_{{\text{obs}},\mathbf{u}} = \{\vb*{s}_{\mathbf{u},i} = (\vb*{x}_{\mathbf{u},i}, t_{\mathbf{u},i})\}_{i=1}^{N_{u}}$. Additionally, $\mathcal{D}_{{\text{obs}}}$ contains $N_{f}$ observations of the source term $f$, denoted $\mathbf{f}_{{\text{obs}}} = \{f_j^{\text{obs}}\}_{j=1}^{N_{f}}$, which are recorded at spatio-temporal locations $\mathbf{S}_{{\text{obs}},\mathbf{f}} = \{\vb*{s}_{\mathbf{f},j} = (\vb*{x}_{\mathbf{f},j}, t_{\mathbf{f},j})\}_{j=1}^{N_{f}}$. These source term observations $f_j^{\text{obs}}$ are assumed to correspond to $f(\vb*{s}_{\mathbf{f},j}; \vb*{\phi}_{\text{true}})$, where $\vb*{\phi}_{\text{true}}$ represents the true value of the PDE parameters that generated the data. Collectively, the observation dataset is thus $\mathcal{D}_{{\text{obs}}} = \{ (\mathbf{S}_{{\text{obs}},\mathbf{u}}, \mathbf{u}_{{\text{obs}}}), (\mathbf{S}_{{\text{obs}},\mathbf{f}}, \mathbf{f}_{{\text{obs}}}) \}$.

\subsection{Two-stage inference framework with a physics-based  DKL surrogate}
\label{sec:two_stage}
A key challenge lies in jointly inferring the unknown PDE parameters $\boldsymbol{\phi}$  (Eq. \eqref{eq:pde_definition_method}) and optimizing the DKL model parameters $(\boldsymbol{\theta}, \boldsymbol{\psi})$. 
The DKL parameters $(\boldsymbol{\theta}, \boldsymbol{\psi})$ are typically learned by minimizing the negative log marginal likelihood (NLML), i.e., $-\log p(\mathcal{D}_{{\text{obs}}} \mid \boldsymbol{\theta}, \boldsymbol{\psi})$. 
The NLML is typically analytically tractable under a Gaussian prior and Gaussian likelihood.
However, in the problem considered here, the likelihood $p(\mathcal{D}_{obs} \mid \boldsymbol{\theta}, \boldsymbol{\psi}, \boldsymbol{\phi})$ also depends on the unknown PDE parameters $\boldsymbol{\phi}$, which is not a Gaussian function w.r.t $\boldsymbol{\phi}$.
Particularly, obtaining the marginal likelihood $p(\mathcal{D}_{{\text{obs}}} \mid \boldsymbol{\theta}, \boldsymbol{\psi})$ for the DKL parameters requires integrating out $\boldsymbol{\phi}$ as follows:
\begin{equation} \label{eq:lml_intractable_integral}
p(\mathcal{D}_{\text{obs}} \mid \boldsymbol{\theta}, \boldsymbol{\psi}) = \int p(\mathcal{D}_{{\text{obs}}} \mid \boldsymbol{\theta}, \boldsymbol{\psi}, \boldsymbol{\phi}) p(\boldsymbol{\phi}) d\boldsymbol{\phi}\,.
\end{equation}
The PDE operator $\mathcal{A}[\cdot;\boldsymbol{\phi}]$ typically introduces a non-Gaussian dependence on $\boldsymbol{\phi}$ within the integrand $p(\mathcal{D}_{{\text{obs}}} \mid \boldsymbol{\theta}, \boldsymbol{\psi}, \boldsymbol{\phi})$. 
As a result, the marginal likelihood $p(\mathcal{D}_{{\text{obs}}} \mid \boldsymbol{\theta}, \boldsymbol{\psi})$ generally lacks the analytical form required for standard NLML-based optimization.
Its direct computation and subsequent minimization with respect to $(\boldsymbol{\theta}, \boldsymbol{\psi})$ thus become intractable.

To address this, we propose a two-stage methodology centred on the physics-based  DKL model as the core solution surrogate, separating NN's parameters optimization from the Bayesian inference of $\boldsymbol{\phi}$.

The first stage, physics-based pretraining (section \ref{sec:stage_1}), is designed to yield an optimized set of neural network parameters $\boldsymbol{\theta}_{\text{fix}}$, along with initial estimates both for the PDE parameters, $\boldsymbol{\phi}_{\text{pre}}$, and for the base kernel hyperparameters, $\boldsymbol{\psi}_{\text{pre}}$. 
These are obtained by minimizing a composite physics-based loss function. 
The second stage (section \ref{sec:stage_2}) then utilizes the DKL model with its neural network component defined by the fixed parameters $\boldsymbol{\theta}_{\text{fix}}$, to perform Bayesian inference jointly for $\boldsymbol{\phi}$ and $\boldsymbol{\psi}$. 
This stage yields the joint posterior distribution $p(\boldsymbol{\phi}, \boldsymbol{\psi} \mid \mathcal{D}_{{\text{obs}}}, \boldsymbol{\theta}_{\text{fix}})$. HMC is then used to draw samples from this joint distribution. The marginal posterior $p(\boldsymbol{\phi} \mid \mathcal{D}_{{\text{obs}}}, \boldsymbol{\theta}_{\text{fix}})$ is subsequently obtained by considering only the $\boldsymbol{\phi}$-components of these joint samples, which marginalizes $\boldsymbol{\psi}$. 
This two-stage approach enhances the tractability of inference and provides robust uncertainty quantification for $\boldsymbol{\phi}$.

\subsubsection{Stage 1: physics-based pretraining for deep kernel}
\label{sec:stage_1}
The initial pretraining stage aims to determine an optimized parameter set $\boldsymbol{\theta}_{\text{fix}}$ for the neural network feature extractor $g_{\boldsymbol{\theta}}$ in the DKL model. It also provides well-informed initial estimates for the DKL base kernel hyperparameters  $\boldsymbol{\psi}_{\text{pre}}$, and for the PDE parameters $\boldsymbol{\phi}_{\text{pre}}$.  These are obtained by minimizing a composite loss $\mathcal{L}(\boldsymbol{\theta}, \boldsymbol{\psi}, \boldsymbol{\phi})$ that jointly accounts for data fidelity, physical consistency with the governing PDE, and GP-based regularization.

The composite loss is formulated as:
\begin{equation}
\mathcal{L}(\boldsymbol{\theta}, \boldsymbol{\psi}, \boldsymbol{\phi}) = w_{\text{data}} \mathcal{L}_{\text{data}}(\boldsymbol{\theta}, \boldsymbol{\psi}) + w_{\text{PDE}} \mathcal{L}_{\text{PDE}}(\boldsymbol{\theta}, \boldsymbol{\psi}, \boldsymbol{\phi}) + w_{\text{GP}} \mathcal{L}_{\text{GP}}(\boldsymbol{\theta}, \boldsymbol{\psi},  \boldsymbol{\phi})\,.
\label{eq:composite_loss_stage1}
\end{equation}
Here, $w_{\text{data}}$, $w_{\text{PDE}}$, and $w_{\text{GP}}$ are  positive scalar weights that balance the contribution of each term. 

The data fidelity term $\mathcal{L}_{\text{data}}$ enforces fidelity to the direct observations of the state variable $\mathbf{u}_{{\text{obs}}}$. It is the mean squared error between $\mathbf{u}_{{\text{obs}}}$ and the DKL model's predictive mean:
\begin{equation}
\mathcal{L}_{\text{data}}(\boldsymbol{\theta}, \boldsymbol{\psi}) = \frac{1}{N_{u}} \sum_{i=1}^{N_{u}}\left( u_i^{\text{obs}} - \mu_{\text{DKL}}(\mathbf{s}_{\mathbf{u},i} \mid \mathbf{S}_{{\text{obs}},\mathbf{u}}, \mathbf{u}_{{\text{obs}}}, \boldsymbol{\theta}, \boldsymbol{\psi}, \tau_u^2) \right)^2.
\label{eq:loss_data_stage1}
\end{equation}
Here, $\mu_{\text{DKL}}(\mathbf{s}_{\mathbf{u},i} \mid \mathbf{S}_{{\text{obs}},\mathbf{u}}, \mathbf{u}_{{\text{obs}}}, \boldsymbol{\theta}, \boldsymbol{\psi}, \tau_u^2)$ is the posterior predictive mean of the DKL surrogate, conditioned on the state observations. Its analytical form is given by the standard GP predictive mean equation (Eq. \eqref{eq:gpr_predictive_mean}), with the kernel replaced by the deep kernel $\kappa_{\text{DKL}}(\cdot, \cdot; \boldsymbol{\theta}, \boldsymbol{\psi})$. The term $\tau_u^2$ represents the variance of the observation noise assumed by the GP model (see Eq. \eqref{eq:gpr_observation_model}). For the case of noiseless data, as considered in this work, $\tau_u^2$ is treated as a "nugget" (e.g., $10^{-6}$) to ensure numerical stability of the kernel matrix inversion required in the computation of $\mu_{{\text{DKL}}}(\cdot)$.

The second term $\mathcal{L}_{\text{PDE}}(\boldsymbol{\theta}, \boldsymbol{\psi}, \boldsymbol{\phi})$, penalizes the residual of the governing PDE  \eqref{eq:pde_operator_form_method} at $N_{\text{col}}$ collocation points $\mathbf{S}_{\text{col}} = \{\mathbf{s}_{\text{col},j}\}_{j=1}^{N_{\text{col}}}$. The DKL predictive mean $\mu_{\text{DKL}}(\mathbf{s} \mid \mathbf{S}_{{\text{obs}},\mathbf{u}}, \mathbf{u}_{{\text{obs}}}, \boldsymbol{\theta}, \boldsymbol{\psi}, \tau_u^2)$ serves as the surrogate for $u(\mathbf{s})$:
\begin{equation}
\mathcal{L}_{\text{PDE}}(\boldsymbol{\theta}, \boldsymbol{\psi}, \boldsymbol{\phi}) = \frac{1}{N_{\text{col}}} \sum_{j=1}^{N_{\text{col}}}\left\| \mathcal{A}[\mu_{\text{DKL}}(\mathbf{s}_{\text{col},j} \mid \mathbf{S}_{{\text{obs}},\mathbf{u}}, \mathbf{u}_{{\text{obs}}}, \boldsymbol{\theta}, \boldsymbol{\psi}, \tau_u^2); \boldsymbol{\phi}] - f(\mathbf{s}_{\text{col},j}; \boldsymbol{\phi}) \right\|^2.
\label{eq:loss_pde_stage1}
\end{equation}
Computation of the term $\mathcal{A}[\cdot; \boldsymbol{\phi}]$ within the residual involves derivatives of $\mu_{\text{DKL}}$ with respect to the coordinates in $\mathbf{s}_{\text{col},j}$.
These derivatives are obtained
using AD, propagating through the entire DKL structure.

The third term $\mathcal{L}_{\text{GP}}(\boldsymbol{\theta}, \boldsymbol{\psi}, \boldsymbol{\phi})$ is the NLML of the full observation dataset $\mathcal{D}_{{\text{obs}}} = \{ (\mathbf{S}_{{\text{obs}},\mathbf{u}}, \mathbf{u}_{{\text{obs}}}), (\mathbf{S}_{{\text{obs}},\mathbf{f}}, \mathbf{f}_{{\text{obs}}}) \}$. 
This NLML is derived from the joint GP model where $u(\mathbf{s}) \sim \mathcal{GP}(0, \kappa_{\text{DKL}}(\cdot,\cdot;\boldsymbol{\theta},\boldsymbol{\psi}))$ and $f(\mathbf{s}) = \mathcal{A}[u(\mathbf{s}); \boldsymbol{\phi}]$.  
Let $\mathbf{d}_{\text{joint}} = [\mathbf{u}_{{\text{obs}}}^T, \mathbf{f}_{{\text{obs}}}^T]^T$ be the concatenated vector of all $N_{\text{total}} = N_u + N_f$ observations.  
To ensure numerical stability during matrix inversion, we introduce a noise covariance matrix, $\mathbf{\Sigma}_{\text{noise}}$. Assuming independent noise for the state and source term observations, it is defined as a diagonal matrix:$\mathbf{\Sigma}_{\text{noise}} = \text{diag}(\tau_u^2 \mathbf{I}_{N_u}, \tau_f^2 \mathbf{I}_{N_f})$, where $\tau_u^2$ and $\tau_f^2$ are the noise variances.
The NLML \eqref{eq:dkl_log_marginal_likelihood} is then:
\begin{equation}
\mathcal{L}_{\text{GP}}(\boldsymbol{\theta}, \boldsymbol{\psi}, \boldsymbol{\phi}) = \frac{1}{2}\mathbf{d}_{\text{joint}}^T (\mathbf{K}_{\text{joint}}(\boldsymbol{\theta}, \boldsymbol{\psi}, \boldsymbol{\phi}) + \mathbf{\Sigma}_{\text{noise}})^{-1} \mathbf{d}_{\text{joint}} + \frac{1}{2}\log|\mathbf{K}_{\text{joint}}(\boldsymbol{\theta}, \boldsymbol{\psi}, \boldsymbol{\phi}) + \mathbf{\Sigma}_{\text{noise}}| + \frac{N_{\text{total}}}{2}\log(2\pi) \,.
\label{eq:loss_gp_joint_nll_stage1}
\end{equation}
The joint covariance matrix $\mathbf{K}_{\text{joint}}(\boldsymbol{\theta}_{\text{fix}}, \boldsymbol{\psi}, \boldsymbol{\phi})$ has the block structure:
\begin{equation} \label{eq:K_joint_structure_stage2}
\mathbf{K}_{\text{joint}} = \begin{bmatrix}
\mathbf{K}_{uu} & \mathbf{K}_{uf} \\
\mathbf{K}_{fu} & \mathbf{K}_{ff}
\end{bmatrix}.
\end{equation}
where each block is a matrix defined by applying the operator $\mathcal{A}[\cdot;\boldsymbol{\phi}]$ to the deep kernel $\kappa_{\text{DKL}}(\cdot,\cdot;\boldsymbol{\theta}_{\text{fix}},\boldsymbol{\psi})$. The blocks are given by:
\begin{align}
\mathbf{K}_{uu} &= \kappa_{\text{DKL}}(\mathbf{S}_{{\text{obs}},\mathbf{u}}, \mathbf{S}_{{\text{obs}},\mathbf{u}}; \boldsymbol{\theta}_{\text{fix}}, \boldsymbol{\psi})\,, \label{eq:Kuu_block_def} \\
\mathbf{K}_{uf} &= \mathcal{A}_{\mathbf{s}'}[\kappa_{\text{DKL}}(\mathbf{S}_{{\text{obs}},\mathbf{u}}, \mathbf{S}_{{\text{obs}},\mathbf{f}}; \boldsymbol{\theta}_{\text{fix}}, \boldsymbol{\psi}); \boldsymbol{\phi}]\,, \label{eq:Kuf_block_def} \\
\mathbf{K}_{fu} &= \mathcal{A}_{\mathbf{s}}[\kappa_{\text{DKL}}(\mathbf{S}_{{\text{obs}},\mathbf{f}}, \mathbf{S}_{{\text{obs}},\mathbf{u}}; \boldsymbol{\theta}_{\text{fix}}, \boldsymbol{\psi}); \boldsymbol{\phi}]\,, \label{eq:Kfu_block_def} \\
\mathbf{K}_{ff} &= \mathcal{A}_{\mathbf{s}}\mathcal{A}_{\mathbf{s}'}[\kappa_{\text{DKL}}(\mathbf{S}_{{\text{obs}},\mathbf{f}}, \mathbf{S}_{{\text{obs}},\mathbf{f}}; \boldsymbol{\theta}_{\text{fix}}, \boldsymbol{\psi}); \boldsymbol{\phi}]\,. \label{eq:Kff_block_def}
\end{align}
Here, $\mathbf{K}_{\text{joint}}(\boldsymbol{\theta}_{\text{fix}}, \boldsymbol{\psi}, \boldsymbol{\phi})$ is the $N_{\text{total}} \times N_{\text{total}}$ joint covariance matrix whose blocks are constructed from $\kappa_{\text{DKL}}$ and its transformations by the operator $\mathcal{A}[\cdot;\boldsymbol{\phi}]$, following the principles detailed in Section \ref{par:pde_constrained_gps}.
$\mathcal{A}_{\mathbf{s}}[\kappa(\mathbf{s}, \mathbf{s}')]$ indicates that the operator $\mathcal{A}$ is applied to the first argument, $\mathbf{s}$, of the kernel function for each entry in the resulting matrix. 

Minimizing $\mathcal{L}_{\text{GP}}$ encourages the 
selection of $(\boldsymbol{\theta}, \boldsymbol{\psi}, \boldsymbol{\phi})$ such that the physics-based  DKL model assigns the highest probability to all observed data. 
It can not only provide a probabilistically coherent explanation for all observations under the physics-based  DKL  model but also serves as a form of regularization, as maximizing the marginal likelihood inherently penalizes excessive model complexity.

The pretraining stage thus solves:
\begin{equation}
(\boldsymbol{\theta}_{\text{pre}}, \boldsymbol{\psi}_{\text{pre}}, \boldsymbol{\phi}_{\text{pre}}) = \arg\min_{\boldsymbol{\theta}, \boldsymbol{\psi}, \boldsymbol{\phi}} \mathcal{L}(\boldsymbol{\theta}, \boldsymbol{\psi}, \boldsymbol{\phi})\,.
\label{eq:pretrain_opt_stage1}
\end{equation}
This high-dimensional, non-convex problem is addressed using stochastic gradient methods, with gradients computed via AD. The resulting parameters $\boldsymbol{\theta}_{\text{pre}}$ are then fixed as $\boldsymbol{\theta}_{\text{fix}}$ for stage 2 (\ref{sec:stage_2}), while $\boldsymbol{\psi}_{\text{pre}}$ and $\boldsymbol{\phi}_{\text{pre}}$ serve as prior mean for the Bayesian inference performed in the next stage. The algorithm~\ref{alg:stage1_pretraining} outlines this procedure.

\begin{algorithm}[!ht]
\caption{Stage 1: physics-based deep kernel pretraining}
\label{alg:stage1_pretraining}
\begin{algorithmic}[1]
\Require 
    Observation data $(\mathbf{S}_{\text{obs,u}}, \mathbf{u}_{\text{obs}})$ and $(\mathbf{S}_{\text{obs,f}}, \mathbf{f}_{\text{obs}})$; 
    collocation points $\mathbf{S}_{\text{col}}$; 
    PDE definition $(\mathcal{A}[\cdot;\boldsymbol{\phi}], f(\cdot;\boldsymbol{\phi}))$ from Eq. \eqref{eq:pde_operator_form_method};
    weights for loss terms $w_{\text{data}}, w_{\text{PDE}}, w_{\text{GP}}$; 
    noise covariance $\mathbf{\Sigma}_{\text{noise}}$; optimizer learning rate $\eta$; maximum iterations $N_{\text{iter}}$.
\Ensure Optimized NN parameters $\boldsymbol{\theta}_{\text{fix}}$; Initial estimates $\boldsymbol{\psi}_{\text{pre}}, \boldsymbol{\phi}_{\text{pre}}$.
\State Initialize parameters $\boldsymbol{\theta}^{(0)}, \boldsymbol{\psi}^{(0)}, \boldsymbol{\phi}^{(0)}$.
\State Let $\mathbf{d}_{\text{joint}} = [\mathbf{u}_{\text{obs}}^T, \mathbf{f}_{\text{obs}}^T]^T$.
\For{$k = 0$ to $N_{\text{iter}} - 1$}
    \State Compute DKL conditional predictive mean $\mu_{\text{DKL}}^{(k)}(\cdot) \equiv \mu_{\text{DKL}}(\cdot \mid \mathbf{S}_{\text{obs,u}}, \mathbf{u}_{\text{obs}}, \boldsymbol{\theta}^{(k)}, \boldsymbol{\psi}^{(k)}, \mathbf{\Sigma}_{\text{obs}})$.
    \Comment{\textcolor{gray}{Based on kernel $\kappa_{\text{DKL}}(\cdot;\boldsymbol{\theta}^{(k)},\boldsymbol{\psi}^{(k)})$ (Eq. \eqref{eq:gpr_predictive_mean})}}
    \State Construct joint covariance $\mathbf{K}_{\text{joint}}^{(k)} \gets \mathbf{K}_{\text{joint}}(\boldsymbol{\theta}^{(k)}, \boldsymbol{\psi}^{(k)}, \boldsymbol{\phi}^{(k)})$.
    \State$\mathcal{L}_{\text{data}}^{(k)} \gets \frac{1}{N_u} \sum_{i=1}^{N_u} (u_i^{\text{obs}} - \mu_{\text{DKL}}^{(k)}(\mathbf{s}_{\mathbf{u},i}))^2$.
    \Comment{\textcolor{gray}{Defined in Eq. \eqref{eq:loss_data_stage1}}}
    \State $\mathcal{L}_{\text{PDE}}^{(k)} \gets \frac{1}{N_{\text{col}}} \sum_{j=1}^{N_{\text{col}}} \| \mathcal{A}[\mu_{\text{DKL}}^{(k)}(\mathbf{s}_{\text{col},j}); \boldsymbol{\phi}^{(k)}] - f(\mathbf{s}_{\text{col},j}; \boldsymbol{\phi}^{(k)}) \|^2$.
    \Comment{\textcolor{gray}{Defined in Eq. \eqref{eq:loss_pde_stage1}; Uses $\kappa_{\text{DKL}}$ and $\mathcal{A}[\cdot;\boldsymbol{\phi}^{(k)}]$ as  (\ref{par:pde_constrained_gps})}}
    \State $\mathcal{L}_{\text{GP}}^{(k)} \gets \frac{1}{2}\mathbf{d}_{\text{joint}}^T (\mathbf{K}_{\text{joint}}^{(k)} + \mathbf{\Sigma}_{\text{obs}})^{-1} \mathbf{d}_{\text{joint}} + \frac{1}{2}\log|\mathbf{K}_{\text{joint}}^{(k)} + \mathbf{\Sigma}_{\text{obs}}| + \frac{N_{\text{total}}}{2}\log(2\pi)$.
    \Comment{\textcolor{gray}{Defined in Eq. \eqref{eq:loss_gp_joint_nll_stage1}}}
    \State $\mathcal{L}^{(k)} \gets w_{\text{data}} \mathcal{L}_{\text{data}}^{(k)} + w_{\text{PDE}} \mathcal{L}_{\text{PDE}}^{(k)} + w_{\text{GP}} \mathcal{L}_{\text{GP}}^{(k)}$.
    \Comment{ \textcolor{gray}{Compute total loss define in Eq. \eqref{eq:composite_loss_stage1}}}
    \State Compute gradients $\nabla_{\boldsymbol{\theta}}\mathcal{L}^{(k)}, \nabla_{\boldsymbol{\psi}}\mathcal{L}^{(k)}, \nabla_{\boldsymbol{\phi}}\mathcal{L}^{(k)}$ using AD.
    \Comment{\textcolor{gray}{Compute gradients and update parameters}}
  \State $\boldsymbol{\theta}^{(k+1)} \gets \boldsymbol{\theta}^{(k)} - \eta \nabla_{\boldsymbol{\theta}}\mathcal{L}^{(k)}$.
  \State $\boldsymbol{\psi}^{(k+1)} \gets \boldsymbol{\psi}^{(k)} - \eta \nabla_{\boldsymbol{\psi}}\mathcal{L}^{(k)}$.
  \State $\boldsymbol{\phi}^{(k+1)} \gets \boldsymbol{\phi}^{(k)} - \eta \nabla_{\boldsymbol{\phi}}\mathcal{L}^{(k)}$.
  
\EndFor
\State Set $\boldsymbol{\theta}_{\text{fix}} \gets \boldsymbol{\theta}^{(N_{\text{iter}})}$; $\boldsymbol{\psi}_{\text{pre}} \gets \boldsymbol{\psi}^{(N_{\text{iter}})}$; $\boldsymbol{\phi}_{\text{pre}} \gets \boldsymbol{\phi}^{(N_{\text{iter}})}$.
\State \Return $(\boldsymbol{\theta}_{\text{fix}}, \boldsymbol{\psi}_{\text{pre}}, \boldsymbol{\phi}_{\text{pre}})$.
\end{algorithmic}
\end{algorithm}


\subsubsection{Stage 2: Bayesian inference for PDE parameters \texorpdfstring{$\vb*{\phi}$}{phi}}
\label{sec:stage_2}
Following the pretraining stage, which yields fixed parameters $\boldsymbol{\theta}_{\text{fix}}$ and initial estimates $(\boldsymbol{\phi}_{\text{pre}}, \boldsymbol{\psi}_{\text{pre}})$, this second stage performs a Bayesian inference. 
The objective is to jointly infer the posterior distribution of the PDE parameters $\boldsymbol{\phi}$ and the base kernel hyperparameters $\boldsymbol{\psi}$, conditioned on the observation dataset $\mathcal{D}_{{\text{obs}}}$ (section \ref{sec:problem_formulation}) and $\boldsymbol{\theta}_{\text{fix}}$ from section \ref{sec:stage_1}.

According to Bayes' theorem, the joint posterior distribution of $(\boldsymbol{\phi}, \boldsymbol{\psi})$ is:
\begin{equation}
p(\boldsymbol{\phi}, \boldsymbol{\psi} \mid \mathbf{u}_{\text{obs}}, \mathbf{f}_{\text{obs}}, \mathbf{S}_{{\text{obs}},\mathbf{u}}, \mathbf{S}_{{\text{obs}},\mathbf{f}}, \boldsymbol{\theta}_{\text{fix}}) \propto p(\mathbf{u}_{\text{obs}}, \mathbf{f}_{\text{obs}} \mid \mathbf{S}_{{\text{obs}},\mathbf{u}}, \mathbf{S}_{{\text{obs}},\mathbf{f}}, \boldsymbol{\phi}, \boldsymbol{\psi}, \boldsymbol{\theta}_{\text{fix}}) p(\boldsymbol{\phi}) p(\boldsymbol{\psi}),
\label{eq:stage2_bayes_theorem}
\end{equation}
assuming prior independence between $\boldsymbol{\phi}$ and $\boldsymbol{\psi}$. The prior $p(\boldsymbol{\phi})$ and $p(\boldsymbol{\psi})$ are informed by the prior means. Specific prior choices are detailed in algorithm \ref{alg:stage1_pretraining}.

The likelihood function, $p(\mathbf{u}_{\text{obs}}, \mathbf{f}_{\text{obs}} \mid \mathbf{S}_{{\text{obs}},\mathbf{u}}, \mathbf{S}_{{\text{obs}},\mathbf{f}}, \boldsymbol{\phi}\,, \boldsymbol{\psi}, \boldsymbol{\theta}_{\text{fix}})$, is derived from the joint GP model. This model is defined by the GP prior for the solution field, $u(\mathbf{s}) \sim \mathcal{GP}(0, \kappa_{\text{DKL}}(\cdot,\cdot;\boldsymbol{\theta}_{\text{fix}},\boldsymbol{\psi}))$ (where $\kappa_{\text{DKL}}$ is from Eq. \eqref{eq:dkl_kernel_definition} with fixed $\boldsymbol{\theta}_{\text{fix}}$), and the PDE $f(\mathbf{s}) = \mathcal{A}[u(\mathbf{s}); \boldsymbol{\phi}]$ \eqref{eq:pde_operator_form_method}. The full observation $\mathbf{d}_{\text{joint}}$ is modelled as:
\begin{equation}
\mathbf{d}_{\text{joint}} \mid \boldsymbol{\phi}, \boldsymbol{\psi}, \boldsymbol{\theta}_{\text{fix}} \sim \mathcal{N}\left(\mathbf{0}, \mathbf{K}_{\text{joint}}(\boldsymbol{\theta}_{\text{fix}}, \boldsymbol{\psi}, \boldsymbol{\phi}) + \mathbf{\Sigma}_{\text{noise}}\right)\,.
\label{eq:stage2_joint_gaussian_model}
\end{equation}
Thus, the negative log-likelihood is given by:
\begin{equation}
-\log p(\mathbf{u}_{\text{obs}}, \mathbf{f}_{\text{obs}} \mid \mathbf{S}_{{\text{obs}},\mathbf{u}}, \mathbf{S}_{{\text{obs}},\mathbf{f}}, \boldsymbol{\phi}\,, \boldsymbol{\psi}, \boldsymbol{\theta}_{\text{fix}}) = \frac{1}{2}\mathbf{d}_{\text{joint}}^T (\mathbf{K}_{\text{joint}} + \mathbf{\Sigma}_{\text{noise}})^{-1} \mathbf{d}_{\text{joint}} + \frac{1}{2}\log|\mathbf{K}_{\text{joint}} + \mathbf{\Sigma}_{\text{noise}}| + \frac{N_{\text{total}}}{2}\log(2\pi)\,.
\label{eq:stage2_log_likelihood}
\end{equation}

\paragraph{Interpretation of the likelihood structure}
\label{par:likelihood_interpretation}
The inference of the PDE parameters $\boldsymbol{\phi}$ is founded upon the structure of the log-likelihood function. As derived in Appendix~\ref{app:likelihood_derivation}, the log-likelihood of the joint observations $(\mathbf{u}_{\text{obs}}, \mathbf{f}_{\text{obs}})$ is given by:
\begin{equation}
  \log p(\mathbf{u}_{\text{obs}}, \mathbf{f}_{\text{obs}} \mid \mathbf{S}_{\text{obs,u}}, \mathbf{S}_{\text{obs,f}}, \boldsymbol{\phi}, \boldsymbol{\psi}_{\text{fix}}, \boldsymbol{\theta}_{\text{fix}}) = -\frac{1}{2} \left[ \|\mathbf{f}_{\text{obs}} - \mathbf{h}(\boldsymbol{\phi})\|_{\mathbf{R}_{ff}(\boldsymbol{\phi})}^2 - \log|\mathbf{R}_{ff}(\boldsymbol{\phi})| \right] + C\,,
\end{equation}
where $C$ is a constant independent of $\boldsymbol{\phi}$. The vector $\mathbf{h}(\boldsymbol{\phi})$ and matrix $\mathbf{R}_{ff}(\boldsymbol{\phi})$ are the posterior mean and precision, respectively, defined as:
\begin{align}
  \mathbf{h}(\boldsymbol{\phi}) & := \mathbf{K}_{fu}(\boldsymbol{\phi}) (\mathbf{K}_{uu} + \tau_u^2\mathbf{I}_{N_u})^{-1} \mathbf{u}_{\text{obs}}\,, \\
  \mathbf{R}_{ff}(\boldsymbol{\phi}) & := \left[(\mathbf{K}_{ff}(\boldsymbol{\phi})+ \tau_f^2\mathbf{I}_{N_f})-\mathbf{K}_{fu}(\boldsymbol{\phi}) (\mathbf{K}_{uu} + \tau_u^2\mathbf{I}_{N_u})^{-1}\mathbf{K}_{uf}(\boldsymbol{\phi})\right]^{-1}\,.
\end{align}
The probability landscape over $\boldsymbol{\phi}$ is shaped by two distinct components, jointly defining the log-likelihood function.

The first component, $\|\mathbf{f}_{\text{obs}} - \mathbf{h}(\boldsymbol{\phi})\|_{\mathbf{R}_{ff}(\boldsymbol{\phi})}^2$, is a \textbf{data-fidelity term}. It quantifies the squared Mahalanobis distance \cite{mahalanobis2018generalized} between the observed source data, $\mathbf{f}_{\text{obs}}$, and its prediction, $\mathbf{h}(\boldsymbol{\phi})$. It is the posterior mean of the source term conditioned on the state observations $\mathbf{u}_{\text{obs}}$, calculated via GP regression. It therefore represents the model's best estimate for the source, given the observed state and the physical laws encoded by $\boldsymbol{\phi}$. The weighting matrix, $\mathbf{R}_{ff}(\boldsymbol{\phi})$, is the posterior precision (inverse covariance), which appropriately scales the prediction error by the model's predictive certainty. Consequently, parameter sets $\boldsymbol{\phi}$ that produce predictions in close agreement with the observations are assigned substantially higher likelihood.

The second component, $-\log|\mathbf{R}_{ff}(\boldsymbol{\phi})|$, is a \textbf{model complexity penalty} that provides a natural instantiation of Ockham's razor. The term $\mathbf{R}_{ff}(\boldsymbol{\phi})^{-1}$ is the conditional posterior covariance matrix, and $\log|\mathbf{R}_{ff}(\boldsymbol{\phi})^{-1}|$ measures the logarithmic volume of the uncertainty ellipsoid associated with the prediction $\mathbf{h}(\boldsymbol{\phi})$. Because this term contributes negatively to the log-likelihood (i.e., as $-\frac{1}{2}\log|\mathbf{R}_{ff}(\boldsymbol{\phi})^{-1}|$), the likelihood function inherently penalizes models that exhibit high posterior uncertainty. It parameterized by $\boldsymbol{\phi}$, that yield a more deterministic and confident mapping from source to system state. Models that fit the data with low predictive certainty are considered less plausible than those achieving comparable fit with higher confidence.

The resulting likelihood structure establishes a robust inferential framework. It assigns the highest probability to PDE parameters $\boldsymbol{\phi}$ that represent a physical model that is not only accurate in its predictions, as enforced by the data-fidelity term, but also confident, as favoured by the complexity penalty.


\subsubsection{Hamiltonian Monte Carlo implementation}
\label{sec:hmc_implementation_details}
To approximate the complex posterior distribution $p(\boldsymbol{\phi}, \boldsymbol{\psi} \mid \mathbf{u}_{\text{obs}}, \mathbf{f}_{\text{obs}}, \mathbf{S}_{{\text{obs}},\mathbf{u}}, \mathbf{S}_{{\text{obs}},\mathbf{f}}, \boldsymbol{\theta}_{\text{fix}})$ Eq. \eqref{eq:stage2_bayes_theorem} and quantify the uncertainties in the PDE parameters $\boldsymbol{\phi}$ and DKL base kernel hyperparameters $\boldsymbol{\psi}$, we employ Hamiltonian Monte Carlo (HMC) \cite{neal2011mcmc}, an advanced method known for its efficiency in navigating such challenging distributions (see section \ref{sec:hmc}).

The HMC algorithm requires the definition of a potential energy function $U(\boldsymbol{\phi}, \boldsymbol{\psi})$, which is the negative log of the unnormalized target posterior density. Based on Eq. \eqref{eq:stage2_bayes_theorem} and Eq. \eqref{eq:stage2_log_likelihood}, the potential energy is:
\begin{equation}
U(\boldsymbol{\phi}, \boldsymbol{\psi}) = - \left[ \log p(\mathbf{u}_{\text{obs}}, \mathbf{f}_{\text{obs}} \mid \mathbf{S}_{{\text{obs}},\mathbf{u}}, \mathbf{S}_{{\text{obs}},\mathbf{f}}, \boldsymbol{\phi}\,, \boldsymbol{\psi}, \boldsymbol{\theta}_{\text{fix}}) + \log p(\boldsymbol{\phi}) + \log p(\boldsymbol{\psi}) \right]\,.
\label{eq:potential_energy_hmc}
\end{equation}
A critical requirement for HMC is the gradient of this potential energy with respect to the parameters being sampled, i.e., $\nabla_{\boldsymbol{\phi}} U(\boldsymbol{\phi}, \boldsymbol{\psi})$ and $\nabla_{\boldsymbol{\psi}} U(\boldsymbol{\phi}, \boldsymbol{\psi})$. These gradients are composed of contributions from the log-likelihood and the log-priors:
\begin{align}
\nabla_{\boldsymbol{\phi}} U &= - \left[ \nabla_{\boldsymbol{\phi}} \log p(\mathbf{u}_{\text{obs}}, \mathbf{f}_{\text{obs}} \mid \mathbf{S}_{{\text{obs}},\mathbf{u}}, \mathbf{S}_{{\text{obs}},\mathbf{f}}, \boldsymbol{\phi}\,, \boldsymbol{\psi}, \boldsymbol{\theta}_{\text{fix}}) + \nabla_{\boldsymbol{\phi}} \log p(\boldsymbol{\phi}) \right]\,, \label{eq:grad_U_phi} \\
\nabla_{\boldsymbol{\psi}} U &= - \left[ \nabla_{\boldsymbol{\psi}} \log p(\mathbf{u}_{\text{obs}}, \mathbf{f}_{\text{obs}} \mid \mathbf{S}_{{\text{obs}},\mathbf{u}}, \mathbf{S}_{{\text{obs}},\mathbf{f}}, \boldsymbol{\phi}\,, \boldsymbol{\psi}, \boldsymbol{\theta}_{\text{fix}}) + \nabla_{\boldsymbol{\psi}} \log p(\boldsymbol{\psi}) \right]\,. \label{eq:grad_U_psi}
\end{align}
The gradients of the log-priors, $\nabla_{\boldsymbol{\phi}} \log p(\boldsymbol{\phi})$ and $\nabla_{\boldsymbol{\psi}} \log p(\boldsymbol{\psi})$, are typically available analytically for common prior distributions with tractable log-densities. The primary challenge lies in computing the gradients of the log-likelihood (Eq. \eqref{eq:stage2_log_likelihood}). These gradients involve differentiating through the matrix inverse and log-determinant terms involving the joint covariance matrix $\mathbf{K}_{\text{joint}}(\boldsymbol{\theta}_{\text{fix}}, \boldsymbol{\psi}, \boldsymbol{\phi}) + \mathbf{\Sigma}_{\text{noise}}$. Given that $\mathbf{K}_{\text{joint}}$ depends on $\boldsymbol{\psi}$ (through $\kappa_{\text{DKL}}$) and $\boldsymbol{\phi}$ (through the application of the PDE operator $\mathcal{A}[\cdot;\boldsymbol{\phi}]$ to $\kappa_{\text{DKL}}$), these complex derivatives are efficiently and accurately computed using AD. AD propagates gradients through the entire construction of the likelihood, including the DKL kernel evaluation and the operator transformations.

The specific HMC sampling procedure, including the choice of mass matrix $M$, leapfrog integrator parameters like step size $\epsilon$ and number of steps $L$, and adaptation strategie, is detailed in Algorithm \ref{alg:hmc}. 

\subsubsection{Parameter uncertainty quantification via marginalization}
\label{sec:parameter_uq}

Upon convergence, the HMC algorithm yields a set of $N_s$ joint samples, $\{(\boldsymbol{\phi}^{(j)}, \boldsymbol{\psi}^{(j)})\}_{j=1}^{N_{s}}$, from the posterior distribution $p(\boldsymbol{\phi}, \boldsymbol{\psi} \mid \mathcal{D}_{\text{obs}}, \boldsymbol{\theta}_{\text{fix}})$. While these joint samples are essential for full model averaging, a primary objective of the inverse problem is to quantify the uncertainty of the unknown PDE parameters $\boldsymbol{\phi}$.

This is achieved by obtaining the marginal posterior distribution for $\boldsymbol{\phi}$. A key advantage of Monte Carlo methods is that this marginalization is performed implicitly. The marginal posterior distribution, formally defined by the integral
\begin{equation}
    p(\boldsymbol{\phi} \mid \mathcal{D}_{\text{obs}}, \boldsymbol{\theta}_{\text{fix}}) = \int p(\boldsymbol{\phi}, \boldsymbol{\psi} \mid \mathcal{D}_{\text{obs}}, \boldsymbol{\theta}_{\text{fix}}) \, d\boldsymbol{\psi}\,,
    \label{eq:marginalization_phi}
\end{equation}
is readily represented by considering only the $\boldsymbol{\phi}$-components of the joint samples. The resulting collection, $\{\boldsymbol{\phi}^{(j)}\}_{j=1}^{N_{s}}$, constitutes a valid set of draws from the marginal posterior $p(\boldsymbol{\phi} \mid \mathcal{D}_{\text{obs}}, \boldsymbol{\theta}_{\text{fix}})$.
These marginal samples provide Monte Carlo estimates for posterior statistics.
Similarly, the posterior covariance matrix is estimated by the sample covariance of $\{\boldsymbol{\phi}^{(j)}\}$.

\subsubsection{Forward prediction via Bayesian model averaging}
\label{sec:forward_prediction_bma}

Upon obtaining $N_{s}$ posterior samples $\{(\boldsymbol{\phi}^{(j)}, \boldsymbol{\psi}^{(j)})\}_{j=1}^{N_{s}}$ representing the joint distribution $p(\boldsymbol{\phi}, \boldsymbol{\psi} \mid \mathcal{D}_{\text{obs}}, \boldsymbol{\theta}_{\text{fix}})$ from HMC, we proceed to predict the PDE solution field, denoted by the vector $\mathbf{u}^* \in \mathbb{R}^{N^*}$, at a new set of unobserved spatio-temporal locations $\mathbf{S}^* = \{\mathbf{s}^*_k\}_{k=1}^{N^*}$. This predictive task is accomplished using Bayesian model averaging (BMA), a principled approach that accounts for the uncertainty in the inferred parameters $(\boldsymbol{\phi}, \boldsymbol{\psi})$ by averaging predictions across their joint posterior distribution.

The predictive distribution for the solution vector $\mathbf{u}^*$ under BMA is formally given by marginalizing the conditional predictive distribution with respect to the posterior of the parameters:
\begin{equation}
    p(\mathbf{u}^* \mid \mathbf{S}^*, \mathcal{D}_{\text{obs}}, \boldsymbol{\theta}_{\text{fix}}) = \iint p(\mathbf{u}^* \mid \mathbf{S}^*, \mathcal{D}_{\text{obs}}, \boldsymbol{\phi}, \boldsymbol{\psi}, \boldsymbol{\theta}_{\text{fix}}) \, p(\boldsymbol{\phi}, \boldsymbol{\psi} \mid \mathcal{D}_{\text{obs}}, \boldsymbol{\theta}_{\text{fix}}) \, d\boldsymbol{\phi} \, d\boldsymbol{\psi}\,.
    \label{eq:bma_predictive_integral}
\end{equation}
The $p(\mathbf{u}^* \mid \mathbf{S}^*, \mathcal{D}_{\text{obs}}, \boldsymbol{\phi}, \boldsymbol{\psi}, \boldsymbol{\theta}_{\text{fix}})$ is the Gaussian predictive distribution derived from the DKL model conditioned on the observation dataset $\mathcal{D}_{\text{obs}}$. For a specific posterior sample $(\boldsymbol{\phi}^{(j)}, \boldsymbol{\psi}^{(j)})$, this distribution is a multivariate Gaussian.
Its mean  and covariance matrix  are computed using the standard GP conditioning equations, where the conditioning set is the observation data $\mathbf{d}_{\text{joint}}$. This requires constructing a joint covariance matrix between the test points and the training points using the kernel $\kappa_{\text{DKL}}(\cdot, \cdot; \boldsymbol{\theta}_{\text{fix}}, \boldsymbol{\psi}^{(j)})$ and the operator $\mathcal{A}[\cdot; \boldsymbol{\phi}^{(j)}]$.

The integral in Eq.~\eqref{eq:bma_predictive_integral} is intractable. However, it can be approximated using a Monte Carlo estimate, leveraging the $N_s$ samples drawn from the posterior via HMC. The predictive mean and covariance of $\mathbf{u}^*$ are given by the law of total expectation and total covariance, respectively:
\begin{align}
    \mathbb{E}[\mathbf{u}^* \mid \mathbf{S}^*, \mathcal{D}_{\text{obs}}, \boldsymbol{\theta}_{\text{fix}}] &= \mathbb{E}_{(\boldsymbol{\phi},\boldsymbol{\psi})|\mathcal{D}_{\text{obs}}} \left[ \mathbb{E}[\mathbf{u}^* \mid \mathbf{S}^*, \mathcal{D}_{\text{obs}}, \boldsymbol{\phi}, \boldsymbol{\psi}] \right] \,, \\
    \text{Cov}[\mathbf{u}^* \mid \mathbf{S}^*, \mathcal{D}_{\text{obs}}, \boldsymbol{\theta}_{\text{fix}}] &= \mathbb{E}_{(\boldsymbol{\phi},\boldsymbol{\psi})|\mathcal{D}_{\text{obs}}} \left[ \text{Var}[\mathbf{u}^* \mid \mathbf{S}^*, \mathcal{D}_{\text{obs}}, \boldsymbol{\phi}, \boldsymbol{\psi}] \right] + \text{Var}_{(\boldsymbol{\phi},\boldsymbol{\psi})|\mathcal{D}_{\text{obs}}} \left[ \mathbb{E}[\mathbf{u}^* \mid \mathbf{S}^*, \mathcal{D}_{\text{obs}}, \boldsymbol{\phi}, \boldsymbol{\psi}] \right].
    \label{eq:predictive_covariance_u_bma_law}
\end{align}
In practice, we approximate these quantities by their sample-based estimators. For each posterior sample $(\boldsymbol{\phi}^{(j)}, \boldsymbol{\psi}^{(j)})$, we first compute the conditional predictive mean and covariance. The final BMA predictive mean is then the sample mean of these conditional means, and the BMA predictive covariance is computed by summing the sample mean of the conditional covariances and the sample covariance of the conditional means.
This BMA procedure yields a probabilistic prediction for the solution field, incorporating the uncertainties propagated from both the inferred base kernel hyperparameters $\boldsymbol{\psi}$ and the PDE parameters $\boldsymbol{\phi}$.

\begin{algorithm}[!ht]
\caption{HMC Sampling for $(\boldsymbol{\phi}, \boldsymbol{\psi})$}
\label{alg:hmc}
\begin{algorithmic}[1]
\Require 
    Initial state $\boldsymbol{\xi}^{(0)} = (\boldsymbol{\phi}^{(0)}, \boldsymbol{\psi}^{(0)})$
    Potential energy function $U(\boldsymbol{\xi})$ and its gradient $\nabla_{\boldsymbol{\xi}} U(\boldsymbol{\xi})$ 
    Number of samples $N_{s}$; Warm-up iterations $N_{\text{warmup}}$;
    Leapfrog steps $L$; Leapfrog step size $\epsilon$; Mass matrix $M$.
\Ensure Posterior samples $\{\boldsymbol{\xi}^{(j)} = (\boldsymbol{\phi}^{(j)}, \boldsymbol{\psi}^{(j)})\}_{j=1}^{N_{s}}$.

\State Initialize $\boldsymbol{\xi} \gets \boldsymbol{\xi}^{(0)}$.
\State Initialize empty list $\textit{samples}$.
\For{$i = 1$ to $N_{s} + N_{\text{warmup}}$}
    \State \Comment{\textcolor{gray}{Resample momentum}}
    \State $\mathbf{p} \sim \mathcal{N}(\mathbf{0}, M)$.
    \State Store current state: $\boldsymbol{\xi}_{\text{curr}} \gets \boldsymbol{\xi}$, $\mathbf{p}_{\text{curr}} \gets \mathbf{p}$.
    \State Compute current Hamiltonian $H_{\text{curr}} = U(\boldsymbol{\xi}_{\text{curr}}) + \frac{1}{2}\mathbf{p}_{\text{curr}}^T M^{-1} \mathbf{p}_{\text{curr}}$. \Comment{\textcolor{gray}{Eq. \eqref{eq:hamiltonian}}}

    \State \Comment{\textcolor{gray}{Leapfrog integration}}
    \State $\mathbf{p} \gets \mathbf{p} - (\epsilon/2) \nabla_{\boldsymbol{\xi}} U(\boldsymbol{\xi}_{\text{curr}})$. \Comment{\textcolor{gray}{Half step for momentum}}
    \For{$l = 1$ to $L$}
        \State $\boldsymbol{\xi} \gets \boldsymbol{\xi} + \epsilon M^{-1}\mathbf{p}$. \Comment{\textcolor{gray}{Full step for position $\boldsymbol{\xi}$}}
        \If{$l < L$}
            \State $\mathbf{p} \gets \mathbf{p} - \epsilon \nabla_{\boldsymbol{\xi}} U(\boldsymbol{\xi})$. \Comment{\textcolor{gray}{Full step for momentum}}
        \EndIf
    \EndFor
    \State $\mathbf{p} \gets \mathbf{p} - (\epsilon/2) \nabla_{\boldsymbol{\xi}} U(\boldsymbol{\xi})$. \Comment{\textcolor{gray}{Final half step for momentum}}
    \State Let proposed state be $\boldsymbol{\xi}_{\text{prop}} \gets \boldsymbol{\xi}$ and $\mathbf{p}_{\text{prop}} \gets \mathbf{p}$.

    \State Compute proposed Hamiltonian $H_{\text{prop}} = U(\boldsymbol{\xi}_{\text{prop}}) + \frac{1}{2}\mathbf{p}_{\text{prop}}^T M^{-1} \mathbf{p}_{\text{prop}}$.
    \State Acceptance ratio $\alpha_{\text{acc}} = \min(1, \exp(H_{\text{curr}} - H_{\text{prop}}))$. \Comment{\textcolor{gray}{Eq. \eqref{eq:hmc_acceptance}}}
    \State Draw $u_{\text{rand}} \sim \text{Uniform}(0,1)$.
    \If{$u_{\text{rand}} < \alpha_{\text{acc}}$}
        \State $\boldsymbol{\xi} \gets \boldsymbol{\xi}_{\text{prop}}$. \Comment{\textcolor{gray}{Accept proposal}}
    \Else
        \State $\boldsymbol{\xi} \gets \boldsymbol{\xi}_{\text{curr}}$. \Comment{\textcolor{gray}{Reject proposal, stay at current state}}
    \EndIf
    
    \If{$i > N_{\text{warmup}}$}
        \State Add $\boldsymbol{\xi}$ to $\textit{samples}$.
    \EndIf
\EndFor
\State \Return $\textit{samples}$.
\end{algorithmic}
\end{algorithm}


\section{Numerical Experiments}
\label{sec:numerical_experiments}
This section demonstrates the efficacy of the proposed physics-based  DKL framework for Bayesian inference for physics-based  parameter estimation problems. We present a series of numerical experiments with low and high PDEs, to validate the method's ability to accurately infer unknown PDE parameters $\boldsymbol{\phi}$ and quantify their associated uncertainties. The inference is conditioned on limited, exact observations of the system's state $u$ (as detailed in Section \ref{sec:problem_formulation}). Each experiment employs the two-stage methodology outlined in Section \ref{sec:method}: (1) physics-based DKL pretraining to obtain an optimized set of DKL neural network parameters $\boldsymbol{\theta}_{\text{fix}}$ and initial point estimates $(\boldsymbol{\psi}_{\text{pre}}, \boldsymbol{\phi}_{\text{pre}})$; followed by (2) HMC sampling of the joint posterior distribution $p(\boldsymbol{\phi}, \boldsymbol{\psi} \mid \mathcal{D}_{{\text{obs}}}, \boldsymbol{\theta}_{\text{fix}})$ to characterize the marginal posterior for $\boldsymbol{\phi}$. Our evaluation focuses on the accuracy of the inferred parameters, the fidelity of the posterior uncertainty quantification, and the framework's performance and scalability in high-dimensional settings.
\subsection{Parameter estimation for 1D heat equation}
\label{subsec:heat_1d}

We begin with a fundamental inverse problem: inferring the constant thermal diffusivity coefficient $\alpha$ in the one-dimensional heat equation. The system is defined on the spatial domain $\Omega = [0, 1]$ and time interval $t \in [0, 1]$ by:
\begin{equation} 
\begin{aligned}
    \partial_t u(x, t) - \alpha \, \partial_{xx} u(x, t) &= f(x, t), && (x, t) \in (0, 1) \times (0, 1]\,, \\
    u(x, t) &= 0, && (x, t) \in \{0, 1\} \times [0, 1]\,, \\ 
    u(x, 0) &= \sin (\pi x), && x \in [0, 1]\,, 
\end{aligned}
\label{eq:heat_1d_pde_numexp}
\end{equation}
where the source term is specified as $f(x, t) = e^{-t} \sin (\pi x) (\alpha_{\text{true}} \pi^2 - 1)$. This selection yields the analytical solution $u(x, t) = e^{-t} \sin (\pi x)$. For generating observation data, we set the true diffusivity $\alpha_{\text{true}} = 1.0$. We acquired $N_{u} = 50$ exact observations by evaluating this analytical solution at spatio-temporal points sampled randomly within $\Omega \times (0, 1]$. The inference task is to recover $\alpha$ using these observations, assuming a uniform prior $p(\alpha) \sim U(0, 2)$. The physics-based pretraining employed $N_{f}=100$ collocation points. We ran a single HMC chain with $1500$ warmup iterations followed by $8500$ sampling iterations. 
\begin{figure}[!ht]
    \centering
    \includegraphics[width=0.5\textwidth]{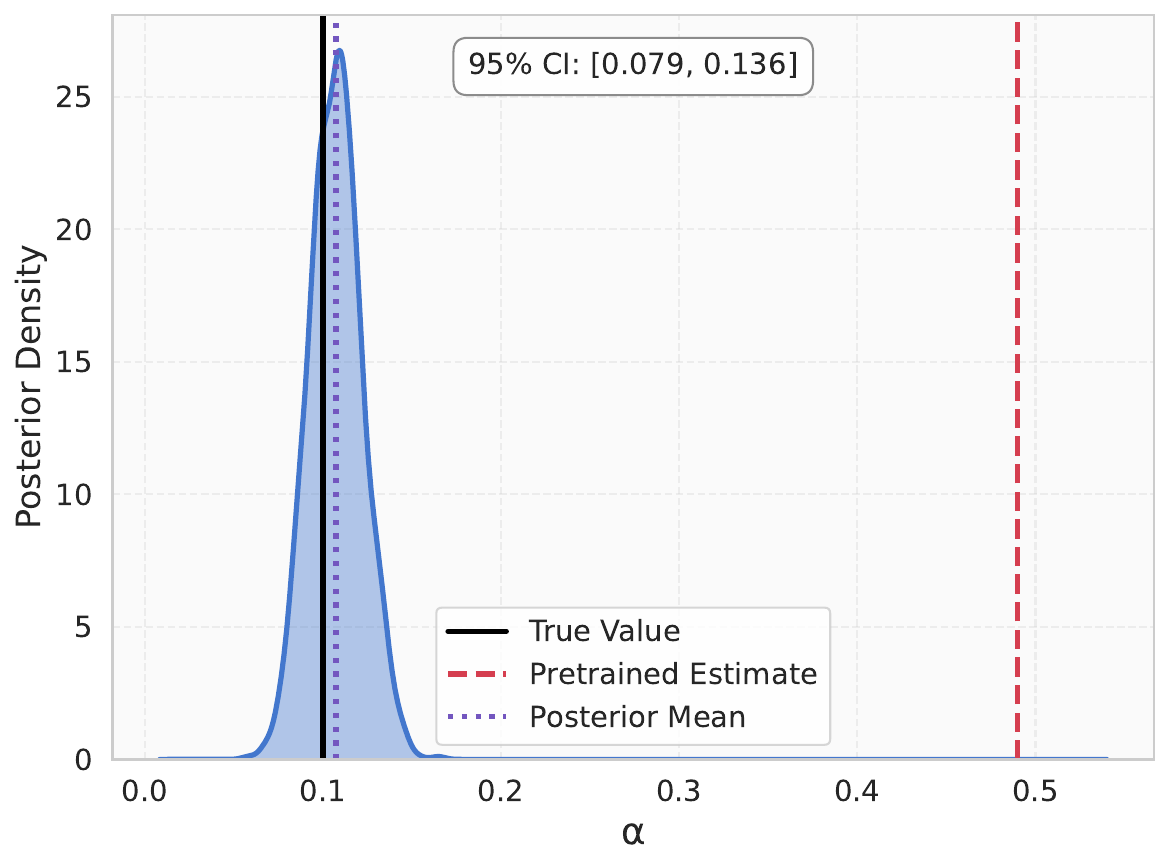}
    \caption{Marginal posterior distribution for the thermal diffusivity $\alpha$ in the 1D heat equation inverse problem (Section~\ref{subsec:heat_1d}). The red dashed line indicates the true value $\alpha_{\text{true}}=1.0$.}
    \label{fig:heat_1d_posterior}
\end{figure}
The resulting marginal posterior distribution for $\alpha$, depicted in Figure~\ref{fig:heat_1d_posterior}, confirms the framework's accuracy. The posterior density is sharply concentrated around the true value $\alpha_{\text{true}} = 1.0$, indicating successful parameter recovery from sparse data. The concentration signifies high confidence in the inferred value, directly attributable to the information assimilated through the physics-based  likelihood during Bayesian inference.

\begin{figure}
    \centering
    \includegraphics[width=1\linewidth]{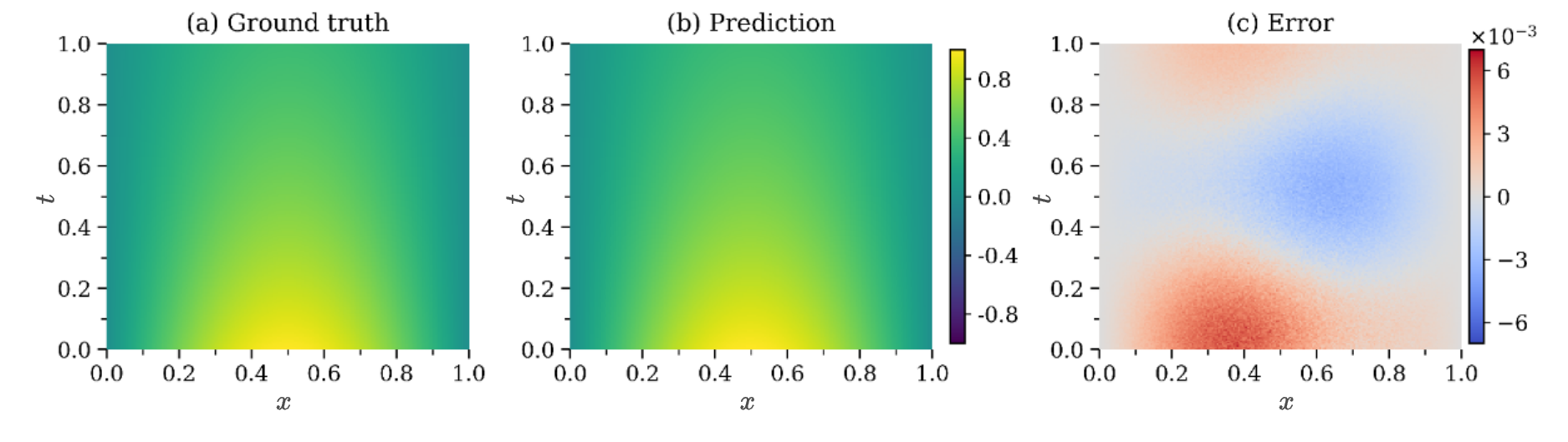}
    \caption{Forward model prediction for the 1D heat equation. (a) The analytical ground truth solution $u(x,t)$. (b) The predictive mean solution obtained via Bayesian model averaging  using the inferred posterior. (c) The point-wise error between the prediction and the ground truth.}
    \label{fig:heat_1d_fwd_pred}
\end{figure}

Beyond achieving an accurate posterior for the parameter $\alpha$, the framework provides a high-fidelity forward predictive model for the solution field $u(x, t)$. This predictive capability is demonstrated in Figure~\ref{fig:heat_1d_fwd_pred}, where the trained DKL surrogate, conditioned on the inferred parameters, is used as a forward solver. The predictive mean of the solution field, obtained via BMA over the posterior samples, shows excellent agreement with the analytical ground truth across the entire spatio-temporal domain. The quantitative accuracy is further highlighted by the point-wise error plot, which reveals that the maximum absolute error is low, on the order of $10^{-3}$. This level of accuracy, achieved using only $N_u=50$ sparse observations, indicates the data-efficiency of our physics-based DKL approach and validates its capability, not only to perform robust parameter inference but also to function as a highly accurate predictive surrogate for the physical system.

\subsection{Parameter estimation  for a 50-dimensional heat equation}
\label{subsec:heat_50d}

To evaluate the framework's performance under the curse of dimensionality, we consider a high-dimensional benchmark for parameter inference based on the heat equation on the hypercube $\Omega=[0,1]^{50}$ with $t\in[0,1]$. The problem is formulated as follows:
\begin{equation}
\begin{aligned}
{\partial_t u(\vb*{x}, t)} - \nabla \cdot \left( \kappa(\vb*{x}; \vb*{\alpha}) \nabla u(\vb*{x}, t) \right) &= f(\vb*{x}, t) \,, && (\vb*{x}, t) \in \Omega \times (0, 1] \,, \\
u(\vb*{x}, t) &= g(\boldsymbol{x},t)\,, && (\vb*{x}, t) \in \partial\Omega \times [0, 1] \,, \\
u(\vb*{x}, 0) &=h(\boldsymbol{x})\,, && \vb*{x} \in \Omega \,.
\end{aligned}
\label{eq:heat_50d_pde_numexp}
\end{equation}
Let $\vb*{\alpha} = (\alpha_1, \alpha_2, \alpha_3)^\top$ denote the parameter vector. The diffusion field is parameterized as:
$
    \kappa(\vb*{x}; \vb*{\alpha}) = 1 + \sum_{k=1}^3 \alpha_k \sin \left(k \pi \|\vb*{x}\|_2 / \sqrt{50} \right)
$.
The true parameters are $\vb*{\alpha}_{\text{true}} = (0.3, 0.1, 0.05)^\top$. Let $\mathbf{1}\in\mathbb{R}^{50}$ denote the all-ones vector. We take an artificial exact solution
$
u_\text{arti}(\boldsymbol{x},t)=\exp(-t)\,\cos\!\left(\frac{1}{50}\,\mathbf{1}^\top \boldsymbol{x}\right).
$
Correspondingly, the initial and boundary conditions are respectively  
$
h(\boldsymbol{x})=u_\text{arti}(\boldsymbol{x},0)=\cos\!\left(\tfrac{1}{50}\,\mathbf{1}^\top \boldsymbol{x}\right) 
$ in $\Omega$, and
$
g(\boldsymbol{x},t)=u_\text{arti}(\boldsymbol{x},t) $ on $\partial\Omega\times(0,1].
$
The right-hand side is defined by $f=\partial_t u_\text{arti}-\nabla\!\cdot(\kappa\nabla u_\text{arti})$, which evaluates to
$
f(\boldsymbol{x},t)
=\exp(-t)\Bigg[
\left(\frac{\kappa}{50}-1\right)
\cos\!\left(\frac{1}{50}\,\mathbf{1}^\top \boldsymbol{x}\right)
 + \frac{1}{50}\,
\left(\sum_{k=1}^3 \alpha_{ k}\,\frac{k\pi}{\sqrt{50}}\,
\cos\!\left(\frac{k\pi}{\sqrt{50}}\,\|\boldsymbol{x}\|_2\right)\right)
\frac{\mathbf{1}^\top \boldsymbol{x}}{\|\boldsymbol{x}\|_2}\,
\sin\!\left(\frac{1}{50}\,\mathbf{1}^\top \boldsymbol{x}\right)
\Bigg]
$.

Simulating or observing data in 50 dimensions poses significant challenges. For this validation, we assume the availability of $N_{u} = 500$ exact observations $\{(\vb*{x}_i, t_i), u_i^{\text{obs}}\}$ notionally drawn from the true underlying process. Independent uniform priors $\alpha_k \sim U(-0.3, 0.3)$ were employed. Reflecting the increased dimensionality, we used $N_{f}=1000$ collocation points. HMC sampling yielded $M=1 \times 10^5$ post-warmup draws in total.
\begin{figure}[!ht]
    \centering
     \includegraphics[width=\textwidth]{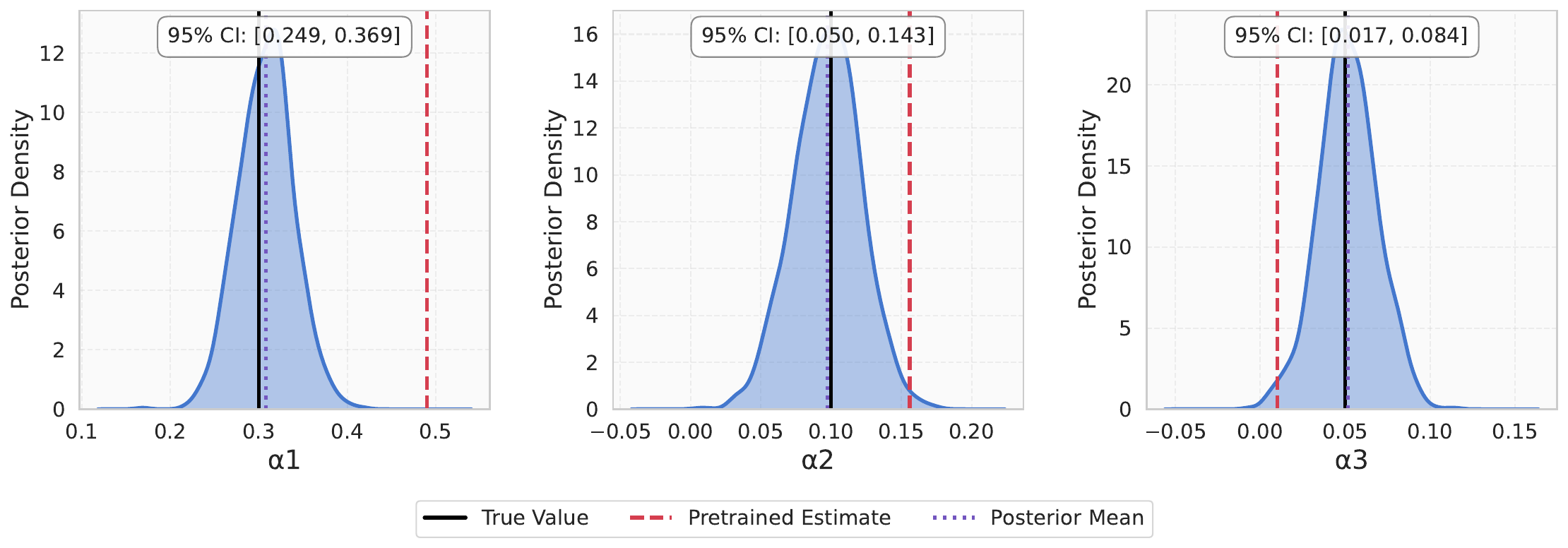}
    \caption{Marginal posterior distributions (estimated via KDE) for the parameters $\alpha_1, \alpha_2, \alpha_3$ in the 50D heat equation inverse problem (Section~\ref{subsec:heat_50d}). Red dashed lines indicate true values.}
    \label{fig:heat_50d_posteriors}
\end{figure}
Figure~\ref{fig:heat_50d_posteriors} displays the resulting marginal posterior distributions for $\alpha_1, \alpha_2, \alpha_3$. Remarkably, even with data that is extremely sparse relative to the volume of the 51-dimensional spatio-temporal domain ($\Omega \times [0,1]$), the proposed approach successfully identifies the region of high posterior probability encompassing the true parameters. The deep kernel's ability to learn a low-dimensional latent representation ($p \ll 51$) from the high-dimensional input $(\vb*{x}, t)$ is crucial for mitigating the curse of dimensionality within the GP framework. While the posterior variances are larger than in the lower-dimensional cases, reflecting the increased uncertainty due to data sparsity, the method correctly pinpoints the plausible parameter ranges, demonstrating its potential applicability to genuinely high-dimensional inverse problems intractable for conventional mesh-based or grid-based approaches.

\subsection{Parameter estimation  for a 50-dimensional advection-diffusion-reaction equation }
\label{subsec:adr_50d}

Finally, we examine the inference of multiple constant physical coefficients in a 50-dimensional advection-diffusion-reaction (ADR) equation, introducing a challenge in terms of simultaneous parameter estimation in a high-dimensional state space. The governing PDE on $\Omega = [-2, 2]^{50}$ for $t \in [0, 1]$ is:
\begin{equation}
\begin{aligned}
\partial_t u(\vb*{x}, t) - \alpha \Delta u(\vb*{x}, t) + \beta\,\mathbf{1}^\top \nabla u(\vb*{x}, t) + \gamma u(\vb*{x}, t) &= f(\vb*{x}, t)\,, && (\vb*{x}, t) \in \Omega \times (0, 1]\,, \\
u(\vb*{x}, t) &= g(\vb*{x},t) \,, && (\vb*{x}, t) \in \partial\Omega \times [0, 1]\,, \\
u(\vb*{x}, 0) &=  h(\vb*{x})\,, && \vb*{x} \in \Omega\,.
\end{aligned}
\label{eq:adr_50d_pde_numexp}
\end{equation}
In this case, the inverse problem refers to inferring the physical parameters $\vb*{\phi} = (\alpha, \beta, \gamma)^\top$, representing the diffusion, advection, and reaction coefficients, respectively. We take an artificial exact solution $u_\text{arti}(\vb*{x}, t)=\exp (-t) \cos \left(\frac{1}{50} \mathbf{1}^{\top} \vb*{x}\right)$. Correspondingly, the initial and boundary conditions are respectively $h(\vb*{x})=u_\text{arti}(\vb*{x}, 0)=\cos \left(\frac{1}{50} \mathbf{1}^{\top} \vb*{x}\right)$ in $\Omega$, and $g(\vb*{x}, t)=u_\text{arti}(\vb*{x}, t)$ on $\partial \Omega \times(0,1]$.  The true values are set to $\vb*{\phi}_{\text{true}} = (0.8, 0.5, 0.2)^\top$, which yields the explicit source term used in this experiment $f(\vb*{x}, t)=\exp (-t)\left[-0.784 \cos \left(\frac{1}{50} \mathbf{1}^{\top} \vb*{x}\right)-0.5 \sin \left(\frac{1}{50} \mathbf{1}^{\top} \vb*{x}\right)\right]$. 

As in the 50D heat equation example, we assume access to $N_{u} = 500$ exact observations. Independent uniform priors were assigned: $\alpha \sim U(0, 2)$, $\beta \sim U(0, 1)$, and $\gamma \sim U(0, 1)$. We utilized $N_{f}=2000$ collocation points, and HMC generated $M=1 \times 10^5$ post-warmup draws.

\begin{figure}[!ht]
    \centering
   \includegraphics[width=\textwidth]{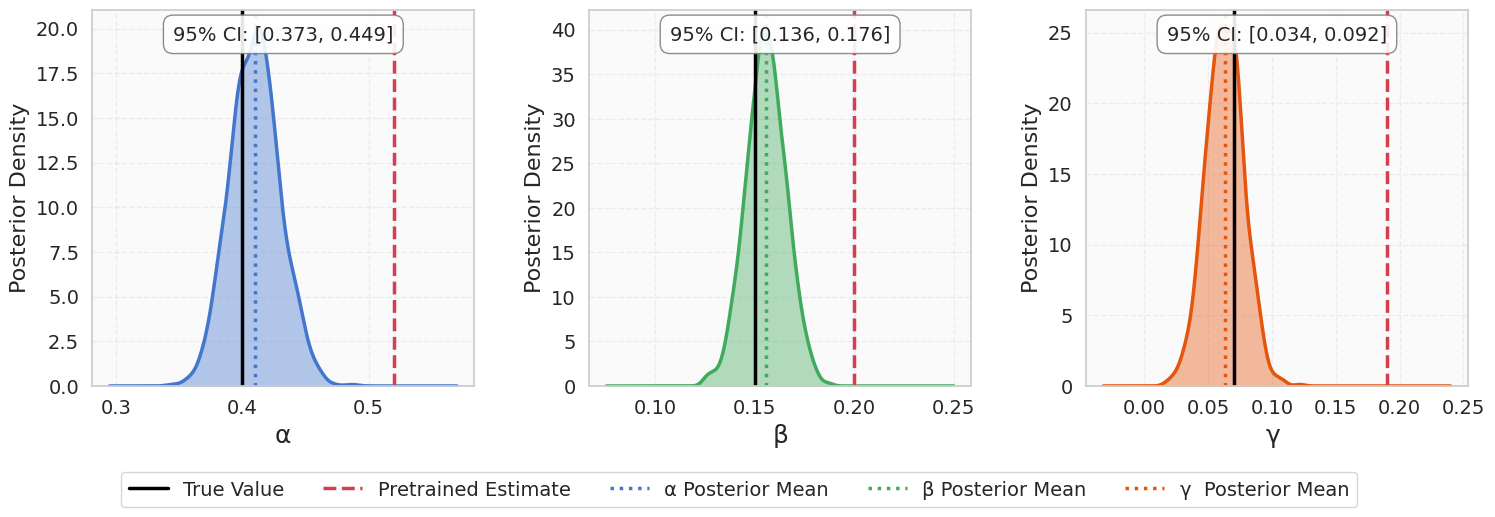} 
    \caption{Marginal posterior distributions for the parameters $\alpha$ (diffusion), $\beta$ (advection), and $\gamma$ (reaction) in the 50D ADR equation inverse problem (Section~\ref{subsec:adr_50d}). Red dashed lines indicate true parameter values. Black dotted lines indicate posterior means. Text boxes show 95\% credible intervals.}
    \label{fig:adr_50d_posteriors}
\end{figure}

The inferred marginal posterior distributions for $\alpha, \beta, \gamma$, presented in Figure~\ref{fig:adr_50d_posteriors}, demonstrate the framework's capacity to accurately estimate multiple constant parameters characterizing a high-dimensional system. Each posterior distribution is clearly unimodal and concentrated around the true parameter value, with the 95\% credible intervals successfully capturing the ground truth. This result underscores the robustness of the physics-based likelihood, derived from the DKL surrogate, in discerning the correct physical parameters governing the high-dimensional dynamics, even from sparse data. The method effectively leverages the underlying PDE structure embedded within the likelihood to disentangle the effects of the different physical processes and identify their corresponding coefficients.

\subsection{Discussion}
\label{sec:discussion}
The numerical experiments provide evidence for the efficacy of our proposed two-stage physics-based  DKL framework in tackling high-dimensional physics-based  inverse problems. A key to this success is the strategic decoupling of the DKL network training (stage 1) from the subsequent Bayesian inference of the PDE parameters $\vb*{\phi}$ and kernel hyperparameters $\boldsymbol{\psi}$ (stage 2). 

Our framework demonstrates the ability to accurately infer parameters and provide robust uncertainty quantification from sparse data, even in challenging high-dimensional settings (up to 50D), highlighting its practical effectiveness. This capability arises from the synergy between DKL’s expressive representation of complex solution fields and the principled incorporation of PDE constraints into the Bayesian likelihood via a joint GP model over $u$ and $\mathcal{A}[u; \boldsymbol{\phi}]$. The successful use of HMC further supports the computational feasibility of our approach for such complex inverse problems.

While our current study has focused on systems governed by linear PDE operators observations, the proposed framework is inherently flexible and extensible. The DKL component is, in principle, expressive enough to approximate solutions of more general PDEs, and the Bayesian inference stage can be readily adapted to handle noisy or incomplete data via appropriate likelihood formulations. 

\section{Conclusion and Outlook}
\label{sec:conclusion_outlook}
In this paper, we have introduced and validated a novel two-stage Bayesian framework that leverages physics-based  DKL for robust parameter inference and uncertainty quantification in high-dimensional PDE inverse problems. Our methodology effectively circumvents the challenges of direct, simultaneous optimization and inference of all DKL and PDE parameters by first employing a physics-based pretraining stage to establish an optimized DKL neural network architecture and sound initial estimates for remaining DKL kernel hyperparameters and PDE parameters. This is followed by a dedicated Bayesian inference stage using HMC to explore the joint posterior of the PDE parameters and DKL kernel hyperparameters, from which the marginal posterior for the target PDE parameters is derived.

The efficacy of this framework was demonstrated through numerical experiments on several challenging parameter estimation, including those in high-dimensional settings. Results consistently showed accurate recovery of PDE parameters from sparse, exact data, alongside reliable quantification of their posterior uncertainties. The proposed two-stage strategy proved instrumental for achieving both computational tractability and high-fidelity inference in these complex scenarios.

Future work will aim to extend this framework's applicability to PDEs involving non-linear operators and to scenarios with noisy observational data, requiring further development in likelihood formulations and potentially more advanced MCMC techniques. Investigating the integration of active learning for optimal data acquisition in high-dimensional spaces also presents a promising research direction. Nevertheless, the presented two-stage physics-based  DKL Bayesian approach offers a powerful and adaptable foundation for tackling a broad spectrum of challenging inverse problems across science and engineering disciplines where dimensionality and rigorous uncertainty assessment are critical.

\section*{Data accessibility}
The data and code supporting the findings of this paper are available on GitHub at the following repository: 
\url{https://github.com/VhaoYan/inversedkl} 
\section*{Acknowledgements}
M.G. acknowledges the financial support from Sectorplan Bèta (the Netherlands) under the focus area \emph{Mathematics of Computational Science} when he held a position at the University of Twente. 
W.Y. acknowledges the China Scholarship Council for its support under No. 202107650017. 
M.G. and C.B. also acknowledge the support from the 4TU Applied Mathematics Institute for the Strategic Research Initiative \emph{Bridging Numerical Analysis and Machine Learning}.

\bibliographystyle{unsrtnat}
\bibliography{references}

\appendix
\section*{Appendix: Derivation of the likelihood function}
\addcontentsline{toc}{section}{Appendix: Derivation of the likelihood function}
\label{app:likelihood_derivation}

\paragraph{Proposition.}
Given the observation dataset $\mathcal{D}_{\text{obs}} = \{ (\mathbf{S}_{\text{obs,u}}, \mathbf{u}_{\text{obs}}), (\mathbf{S}_{\text{obs,f}}, \mathbf{f}_{\text{obs}}) \}$, and for fixed parameters $\boldsymbol{\theta}_{\text{fix}}$ and $\boldsymbol{\psi}_{\text{fix}}$, the log-likelihood of the observations conditioned on their locations and the model parameters can be expressed as
\begin{equation}
  \log p(\mathbf{u}_{\text{obs}}, \mathbf{f}_{\text{obs}} \mid \mathbf{S}_{\text{obs,u}}, \mathbf{S}_{\text{obs,f}}, \boldsymbol{\phi}, \boldsymbol{\psi}_{\text{fix}}, \boldsymbol{\theta}_{\text{fix}}) = -\frac{1}{2} \left[ \|\mathbf{f}_{\text{obs}} - \mathbf{h}(\boldsymbol{\phi})\|_{\mathbf{R}_{ff}(\boldsymbol{\phi})}^2 - \log|\mathbf{R}_{ff}(\boldsymbol{\phi})| \right] + C\,,
\end{equation}
where $C$ is a constant independent of $\boldsymbol{\phi}$. The vector $\mathbf{h}(\boldsymbol{\phi})$ and matrix $\mathbf{R}_{ff}(\boldsymbol{\phi})$ are defined as:
\begin{align}
  \mathbf{h}(\boldsymbol{\phi}) & := \mathbf{K}_{fu}(\boldsymbol{\phi}) (\mathbf{K}_{uu} + \tau_u^2\mathbf{I}_{N_u})^{-1} \mathbf{u}_{\text{obs}}\,, \\
  \mathbf{R}_{ff}(\boldsymbol{\phi}) & := \left[(\mathbf{K}_{ff}(\boldsymbol{\phi})+ \tau_f^2\mathbf{I}_{N_f})-\mathbf{K}_{fu}(\boldsymbol{\phi}) (\mathbf{K}_{uu} + \tau_u^2\mathbf{I}_{N_u})^{-1}\mathbf{K}_{uf}(\boldsymbol{\phi})\right]^{-1}\,.
\end{align}

\begin{proof}
Let the joint observation vector be $\mathbf{d}_{\text{joint}} = [\mathbf{u}_{\text{obs}}^T, \mathbf{f}_{\text{obs}}^T]^T$. Under the GP assumption, the log-likelihood of the observations, conditioned on their locations and the model parameters, is given by the full expression for a multivariate normal distribution:
\begin{equation}
  \log p(\mathbf{u}_{\text{obs}}, \mathbf{f}_{\text{obs}} \mid \mathbf{S}_{\text{obs,u}}, \mathbf{S}_{\text{obs,f}}, \boldsymbol{\phi}, \boldsymbol{\psi}_{\text{fix}}, \boldsymbol{\theta}_{\text{fix}}) = -\frac{1}{2} \mathbf{d}_{\text{joint}}^T \mathbf{K}_{\text{total}}(\boldsymbol{\phi})^{-1} \mathbf{d}_{\text{joint}} - \frac{1}{2} \log |\mathbf{K}_{\text{total}}(\boldsymbol{\phi})| - \frac{N_u + N_f}{2} \log(2\pi)\,,
\end{equation}
where the total covariance matrix $\mathbf{K}_{\text{total}}(\boldsymbol{\phi})$ is
\begin{equation}
  \mathbf{K}_{\text{total}}(\boldsymbol{\phi}) = 
  \begin{bmatrix}
    \mathbf{K}_{uu}^{\tau} & \mathbf{K}_{uf}(\boldsymbol{\phi}) \\
    \mathbf{K}_{fu}(\boldsymbol{\phi}) & \mathbf{K}_{ff}^{\tau}(\boldsymbol{\phi})
  \end{bmatrix},
\end{equation}
with $\mathbf{K}_{uu}^{\tau} = \mathbf{K}_{uu} + \tau_u^2\mathbf{I}_{N_u}$ and $\mathbf{K}_{ff}^{\tau}(\boldsymbol{\phi}) = \mathbf{K}_{ff}(\boldsymbol{\phi}) + \tau_f^2\mathbf{I}_{N_f}$. Note that $\mathbf{K}_{uu}^{\tau}$ is independent of $\boldsymbol{\phi}$. The proof proceeds by decomposing the quadratic form and the log-determinant term.

\subparagraph{1. Decomposition of the log-determinant.}
Using the block matrix determinant identity, the determinant of $\mathbf{K}_{\text{total}}(\boldsymbol{\phi})$ is
\begin{equation}
  |\mathbf{K}_{\text{total}}(\boldsymbol{\phi})| = |\mathbf{K}_{uu}^{\tau}| \left| \mathbf{K}_{ff}^{\tau}(\boldsymbol{\phi}) - \mathbf{K}_{fu}(\boldsymbol{\phi}) (\mathbf{K}_{uu}^{\tau})^{-1} \mathbf{K}_{uf}(\boldsymbol{\phi}) \right|\,.
\end{equation}
Recalling the definition of $\mathbf{R}_{ff}(\boldsymbol{\phi})$ as the inverse of the Schur complement: 
$$\mathbf{R}_{ff}(\boldsymbol{\phi})^{-1} = \mathbf{K}_{ff}^{\tau}(\boldsymbol{\phi}) - \mathbf{K}_{fu}(\boldsymbol{\phi}) (\mathbf{K}_{uu}^{\tau})^{-1} \mathbf{K}_{uf}(\boldsymbol{\phi})\, .$$
The log-determinant can then be expressed in terms of $\mathbf{R}_{ff}(\boldsymbol{\phi})$:
\begin{equation}
  \log|\mathbf{K}_{\text{total}}(\boldsymbol{\phi})| = \log|\mathbf{K}_{uu}^{\tau}| + \log|\mathbf{R}_{ff}(\boldsymbol{\phi})^{-1}| = \log|\mathbf{K}_{uu}^{\tau}| - \log|\mathbf{R}_{ff}(\boldsymbol{\phi})|\,.
\end{equation}
The term $\log|\mathbf{K}_{uu}^{\tau}|$ is a constant with respect to $\boldsymbol{\phi}$.

\subparagraph{2. Decomposition of the quadratic form $Q(\boldsymbol{\phi})$.}
Let the quadratic form be $Q(\boldsymbol{\phi}) = \mathbf{d}_{\text{joint}}^T \mathbf{K}_{\text{total}}(\boldsymbol{\phi})^{-1} \mathbf{d}_{\text{joint}}$. Using the block matrix inversion formula, we have
\begin{equation}
  \mathbf{K}_{\text{total}}^{-1} = 
  \begin{bmatrix}
    (\mathbf{K}_{uu}^{\tau})^{-1} + (\mathbf{K}_{uu}^{\tau})^{-1} \mathbf{K}_{uf} \mathbf{R}_{ff}(\boldsymbol{\phi}) \mathbf{K}_{fu} (\mathbf{K}_{uu}^{\tau})^{-1} & -(\mathbf{K}_{uu}^{\tau})^{-1} \mathbf{K}_{uf} \mathbf{R}_{ff}(\boldsymbol{\phi}) \\
    -\mathbf{R}_{ff}(\boldsymbol{\phi}) \mathbf{K}_{fu} (\mathbf{K}_{uu}^{\tau})^{-1} & \mathbf{R}_{ff}(\boldsymbol{\phi})
  \end{bmatrix}\,.
\end{equation}
Expanding the quadratic form and completing the square with respect to $\mathbf{f}_{\text{obs}}$ yields:
\begin{equation}
  Q(\boldsymbol{\phi}) = \left( \mathbf{f}_{\text{obs}} - \mathbf{K}_{fu}(\boldsymbol{\phi}) (\mathbf{K}_{uu}^{\tau})^{-1} \mathbf{u}_{\text{obs}} \right)^T \mathbf{R}_{ff}(\boldsymbol{\phi})\left( \mathbf{f}_{\text{obs}} - \mathbf{K}_{fu}(\boldsymbol{\phi}) (\mathbf{K}_{uu}^{\tau})^{-1} \mathbf{u}_{\text{obs}} \right) + \mathbf{u}_{\text{obs}}^T (\mathbf{K}_{uu}^{\tau})^{-1} \mathbf{u}_{\text{obs}}\,.
\end{equation}
Using the definitions of $\mathbf{h}(\boldsymbol{\phi})$ and $\mathbf{R}_{ff}(\boldsymbol{\phi})$, this simplifies to:
\begin{equation}
  Q(\boldsymbol{\phi}) = \|\mathbf{f}_{\text{obs}} - \mathbf{h}(\boldsymbol{\phi})\|_{\mathbf{R}_{ff}(\boldsymbol{\phi})}^2 + \mathbf{u}_{\text{obs}}^T (\mathbf{K}_{uu}^{\tau})^{-1} \mathbf{u}_{\text{obs}}\,.
\end{equation}
The second term, $\mathbf{u}_{\text{obs}}^T (\mathbf{K}_{uu}^{\tau})^{-1} \mathbf{u}_{\text{obs}}$, is independent of $\boldsymbol{\phi}$.

\subparagraph{3. Reconstructing the log-likelihood.}
We substitute the decomposed forms of $Q(\boldsymbol{\phi})$ and $\log|\mathbf{K}_{\text{total}}(\boldsymbol{\phi})|$ back into the full log-likelihood expression:
\begin{align}
  \log p(\mathbf{u}_{\text{obs}}, \mathbf{f}_{\text{obs}} \mid \mathbf{S}_{\text{obs,u}}, \mathbf{S}_{\text{obs,f}}, \boldsymbol{\phi}, \boldsymbol{\psi}_{\text{fix}}, \boldsymbol{\theta}_{\text{fix}}) = \; & -\frac{1}{2} \left[ \|\mathbf{f}_{\text{obs}} - \mathbf{h}(\boldsymbol{\phi})\|_{\mathbf{R}_{ff}(\boldsymbol{\phi})}^2 + \mathbf{u}_{\text{obs}}^T (\mathbf{K}_{uu}^{\tau})^{-1} \mathbf{u}_{\text{obs}} \right] \nonumber \\
  & - \frac{1}{2} \left[ \log|\mathbf{K}_{uu}^{\tau}| - \log|\mathbf{R}_{ff}(\boldsymbol{\phi})| \right] - \frac{N_u + N_f}{2} \log(2\pi)\,.
\end{align}
Finally, we group all terms that are constant with respect to $\boldsymbol{\phi}$ into a single constant $C$:
\begin{equation}
  C := -\frac{1}{2} \mathbf{u}_{\text{obs}}^T (\mathbf{K}_{uu}^{\tau})^{-1} \mathbf{u}_{\text{obs}} - \frac{1}{2} \log|\mathbf{K}_{uu}^{\tau}| - \frac{N_u + N_f}{2} \log(2\pi)\,.
\end{equation}
This yields the final expression for the log-likelihood as a function of $\boldsymbol{\phi}$:
\begin{equation}
  \log p(\mathbf{u}_{\text{obs}}, \mathbf{f}_{\text{obs}} \mid \mathbf{S}_{\text{obs,u}}, \mathbf{S}_{\text{obs,f}}, \boldsymbol{\phi}, \boldsymbol{\psi}_{\text{fix}}, \boldsymbol{\theta}_{\text{fix}}) = -\frac{1}{2} \left[ \|\mathbf{f}_{\text{obs}} - \mathbf{h}(\boldsymbol{\phi})\|_{\mathbf{R}_{ff}(\boldsymbol{\phi})}^2 - \log|\mathbf{R}_{ff}(\boldsymbol{\phi})| \right] + C\,.
\end{equation}
This completes the proof.
\end{proof}

\paragraph{Remark on approximation for MAP estimation.}
Given a Gaussian prior $p(\boldsymbol{\phi}) = \mathcal{N}(\boldsymbol{\phi}; \boldsymbol{\phi}_{\text{pre}}, \mathbf{\Sigma}_{\text{prior}})$. The maximum a posteriori (MAP) estimate is obtained by maximizing the log-posterior distribution with respect to $\boldsymbol{\phi}$. This is equivalent to minimizing the negative of the log-posterior that depend on $\boldsymbol{\phi}$ (if $\boldsymbol{\psi}$ is determined by pretraining):
\begin{equation}
    \boldsymbol{\phi}_{\text{MAP}} = \arg\min_{\boldsymbol{\phi}} \left( \|\mathbf{f}_{\text{obs}} - \mathbf{h}(\boldsymbol{\phi})\|_{\mathbf{R}_{ff}(\boldsymbol{\phi})}^2 - \log|\mathbf{R}_{ff}(\boldsymbol{\phi})| + \|\boldsymbol{\phi} - \boldsymbol{\phi}_{\text{pre}}\|_{\mathbf{\Sigma}_{\text{prior}}^{-1}}^2 \right).
\end{equation}

\end{document}